\documentclass[runningheads]{llncs}
\usepackage[T2A]{fontenc}
\usepackage{amsmath}
\usepackage{amssymb}
\usepackage{amsfonts}

\usepackage{amsthm}
\usepackage[inline]{enumitem}
\usepackage{graphicx}
\usepackage{upgreek}
\usepackage{tikz}
\usetikzlibrary{arrows}
\usepackage[hidelinks,colorlinks=true,allcolors=blue]{hyperref}
\usepackage[title]{appendix}
\usepackage{wrapfig}
\usepackage{framed}

\newcommand{\SCI}{\mathsf{SCI}}
\newcommand{\TSCI}{\mathsf{T}_\SCI}

\tikzset{
	treenode/.style = {draw=none,fill=none,align=center},
	root/.style     = {treenode},
	env/.style      = {treenode},
	dummy/.style    = {treenode}
}

\newcommand{\pv}[1]{\textsf{#1}}

\makeatletter
\newtheorem*{rep@theorem}{\rep@title}
\newcommand{\newreptheorem}[2]{%
	\newenvironment{rep#1}[1]{%
		\def\rep@title{#2 \ref{##1}}%
		\begin{rep@theorem}}%
		{\end{rep@theorem}}}
\makeatother

\newreptheorem{proposition}{Proposition}

\begin{document}
\title{Tableau-based decision procedure for non-Fregean logic of sentential identity\thanks{Research reported in this paper is supported by the National Science Centre, Poland (grant number: UMO-2017/25/B/HS1/00503).}}
%
%
\author{Joanna Golińska-Pilarek\inst{1}\orcidID{0000-0001-8546-2615} \and
Taneli Huuskonen\inst{1}\orcidID{0000-0001-7882-8236} \and
Michał Zawidzki\inst{2,3}\orcidID{0000-0003-2909-5923}}
\authorrunning{J. Golińska-Pilarek et al.}
%
\institute{Faculty of Philosophy, University of Warsaw, 3 Krakowskie Przedmiescie St. 00-927 Warsaw, Poland \and
Department of Computer Science, University of Oxford, Oxford OX1 3QD, UK \and
Department of Logic, University of Lodz, 3/5 Lindleya St., 90-131 Łódź, Poland
\\
\email{j.golinska@uw.edu.pl \\ taneli@poczta.onet.pl \\ michal.zawidzki@cs.ox.ac.uk}}
\maketitle              
\begin{abstract}
Sentential Calculus with Identity ($\SCI$) is an extension of classical propositional logic, featuring a new connective of identity between formulas. In $\SCI$ two formulas are said to be identical if they share the same denotation. In the semantics of the logic, truth values are distinguished from denotations, hence the identity connective is strictly stronger than classical equivalence. In this paper we present a sound, complete, and terminating algorithm deciding the satisfiability of $\SCI$-formulas, based on labelled tableaux. To the best of our knowledge, it is the first implemented decision procedure for $\SCI$ which runs in \textsc{NP}, i.e., is complexity-optimal. The obtained complexity bound is a result of dividing derivation rules in the algorithm into two sets: \emph{decomposition} and \emph{equality} rules, whose interplay yields derivation trees with branches of polynomial length with respect to the size of the investigated formula. We describe an implementation of the procedure and compare its performance with implementations of other calculi for $\SCI$ (for which, however, the termination results were not established). We show possible refinements of our algorithm and discuss the possibility of extending it to other non-Fregean logics.

\end{abstract}

\section{Introduction}\label{sect::Introduction}

In this paper, we present a decision procedure for the non-Fregean sentential calculus with identity $\SCI$. The contribution of the paper is twofold. First of all, this is the first implemented and complexity-optimal decision procedure for $\SCI$, although several deduction systems for $\SCI$ have already been presented in the literature.\label{footnote::1} Second, our decision procedure is constructed in the paradigm of labelled tableaux, which makes the whole approach more robust to modifications and extensions to other non-Fregean logics.

Non-Fregean logic is an alternative to both classical and many non-classical systems whose semantics identifies semantical correlates of sentences with their logical values. According to the classical approach in model theory, semantical structures (realities) correspond to the language that is meant to describe them, and therefore, symbols and expressions of that language, such as individual constants or relational symbols, have their denotations in these structures (respectively, objects or relations between objects). However, sentences are treated differently, as they are interpreted in models only in terms of logical values or other semantical relations such as satisfaction or truth. This classical approach allows us to answer the very basic logical question of whether the sentences are logically equivalent; however, it does not provide any tool that would allow to check whether the sentences describe or refer to the same situation, or have the same meaning. Thus, the main motivation for non-Fregean logic was the need for an extensional and two-valued logic that could be used to represent semantical denotations of sentences that -- depending on the underlying philosophical theory of language or the reality to which a logic is supposed to refer -- could be understood as situations, states of affairs, meanings, etc. In order to express (non)identities or other interactions between the referents of sentences, at least the universe of denotations of sentences needs to be added to the semantics and the new \emph{identity} connective to the language.
     
The minimal two-valued non-Fregean propositional logic $\SCI$ (\emph{Sentential Calculus with Identity}), introduced by Suszko (see~\cite{sus75}), is an extension of classical propositional logic with a new binary connective of identity ($\equiv$) and axioms reflecting its fundamental properties. The identity connective represents the identity of the denotations of sentences, and so, an expression `$\upvarphi \equiv \uppsi$' should be read as `the sentences $\upvarphi$ and $\uppsi$ describe the same \guillemotleft{}thing\guillemotright'. The semantics for $\SCI$ is based on structures determined by a universe of the denotations of sentences, a set of facts (those denotations that actually hold), and operations corresponding to all the connectives. The identity connective is then interpreted as an operation representing an equivalence relation that additionally satisfies the extensionality property. In the non-Fregean approach the identity and equivalence connectives are in general not equivalent: two sentences with the same truth value can have different denotations. Take, for instance, the following three statements:
\begin{enumerate}[label=\Alph*]
	\item `There is an effective method for determining whether an arbitrary formula of classical propositional logic is a theorem of that logic.'
	\item `Classical propositional logic is finitely axiomatizable, has a recursive set of recursive rules and enjoys the finite model property.'
	\item `Classical propositional logic is Post consistent.'
\end{enumerate}
A, B, C are all (necessarily) true as theorems of mathematical logic. Therefore, they are pairwise logically equivalent, that is, all three equivalences: $\text{A}\leftrightarrow\text{B}$, $\text{B}\leftrightarrow\text{C}$, and $\text{A}\leftrightarrow\text{C}$ hold. One can fairly claim that A and B refer to the same fact, so $\text{A}\equiv\text{B}$, but C has clearly a different semantic correlate than both A and B, as decidability is independent of Post consistency. Thus, we have $\text{A}\not\equiv\text{C}$ and $\text{B}\not\equiv\text{C}$.

It is known that the class of all non-equivalent non-Fregean propositional logics satisfying the laws of classical logic is uncountable~\cite{gohu}, and some of these logics are equivalent to the well-known non-classical logics (e.g., modal logics $\mathsf{S}4$ and $\mathsf{S}5$, many-valued logics). Higher-order non-Fregean logics are very expressive. In particular, a logic obtained from $\SCI$ by adding propositional quantifiers is undecidable and can express many mathematical theories, e.g., Peano arithmetic, the theory of groups, rings, and fields~\cite{gh16}. Furthermore, non-classical and deviant modifications of $\SCI$ have been developed and extensively studied in the literature, in particular intuitionistic logics~\cite{Lukowski1990b,lew09,ChLJ19}, modal and epistemic logics \cite{lew11,lew15}, logics with non-classical identity~\cite{ish}, paraconsistent~\cite{gol16,gohu18}. The non-Fregean approach could turn out to be more adequate than the classical one in cognitive science or natural language processing. Moreover, non-Fregean logic could serve as a general framework for comparing different aspects of logics with incompatible languages and semantics and help in addressing the question of which class of logics handles logical symbols in the most adequate way from the perspective of natural language.

In the original works by Suszko and Bloom the deduction system for $\SCI$ was defined in the Hilbert style~\cite{BloomSuszko1971,bloomSuszko1972}. Sound and complete deduction systems which are better suited for automated theorem proving were constructed later: Gentzen sequent calculi \cite{Michaels1974,Wasilewska1976,Wasilewska1984,chlebowski} and dual tableau systems \cite{jgp07,org11,jgpwelle}. A detailed presentation of all of them can be found in \cite{jgpwelle}. The main disadvantage of the aforementioned systems is that they are not decision procedures, while $\SCI$ is decidable and in particular in \textsc{NP}~\cite[Theorem 2.3]{bloomSuszko1972}. Although the system by Wasilewska~\cite{Wasilewska1976} can be seen as a meta-tool for deciding validity of $\SCI$-formulas, it is equipped with external meta-machinery that is not a part of the system itself. As a result, it constitutes another proof for decidability of $\SCI$, rather than being a decision procedure in the classical sense of the term, that is suitable for computer implementations. In~\cite{GPZ19} a tableau-based algorithm for $\SCI$ was presented as a work-in-progress. The decision procedure presented in this paper is a result of a substantial remodelling of the preliminary system introduced in~\cite{GPZ19}, for which we prove soundness and completeness, present surprisingly straightforward proofs of termination and membership in \textsc{NP}, and provide an implementation.

In this paper, we present a new deduction system $\TSCI$ for the logic $\SCI$,
based on labelled tableaux. To the best of our knowledge, it is the first
decision procedure for $\SCI$. Moreover, its upper complexity bound, that is \textsc{NP}, matches the complexity class of the satisfiability problem for $\SCI$,
thus, making the algorithm complexity-optimal. $\TSCI$ is built in the paradigm of labelled tableaux. The language of deduction is an extension of the $\SCI$-language with two sorts of labels representing the denotations of formulas (i.e., \guillemotleft{}facts\guillemotright{} and \guillemotleft{}non-facts\guillemotright) as well as with the equality and the inequality relation that can hold between labels. (In)Equality formulas occurring in a derivation tree provide additional information on identity or distinctness of the denotations of formulas. In Section~\ref{sect::SCI}, we provide a formal overview of the logic ${\sf SCI}$, in Section~\ref{sect::tableaux}, we introduce the tableau algorithm $\TSCI$ and prove its soundness, completeness, and termination, 
establish that it is complexity-optimal with respect to $\SCI$-satisfiability,
and show a possible refinement thereof. In Section~\ref{sect::implementation}, we discuss an
implementation of $\TSCI$ and compare it with an older prover based on a heuristic,
unproven algorithm. Conclusions and directions of further research are presented in Section~\ref{sect::Conclusions}.

\section{$\SCI$}\label{sect::SCI}
\paragraph{Syntax}
Let $ \mathcal{L}_{\SCI}$ be a language of the logic $\SCI$ with the alphabet $ \langle \mathsf{AF},\neg,\to,\equiv\rangle $, where $ \mathsf{AF}=\{\pv{p},\pv{q},\pv{r},\ldots\} $ is a denumerable set of \emph{atomic formulas}. The set $\mathsf{FOR}$ of \emph{$\SCI$-formulas} is defined by the following abstract grammar:
\begin{align*}
\upvarphi::=\pv{p}\mid\neg\upvarphi\mid\upvarphi\to\upvarphi\mid\upvarphi\equiv\upvarphi,\end{align*}
where $\pv{p}\in\mathsf{AF}$. 

\paragraph{Axiomatization} The logic $\SCI$ is axiomatized by the following set of truth-functional (\ref{ax::1}--\ref{ax::3}) and identity (\ref{ax::4}--\ref{ax::8}) axiom schemes:
\begin{enumerate}
\item\label{ax::1} $\upvarphi\to(\uppsi\to\upvarphi)$
\item\label{ax::2} $(\upvarphi\to(\uppsi\to\upchi))\to((\upvarphi\to\uppsi)\to(\upvarphi\to\upchi))$
\item\label{ax::3} $(\neg\upvarphi\to\neg\uppsi)\to(\uppsi\to\upvarphi)$
\item\label{ax::4} $\upvarphi\equiv\upvarphi$
\item\label{ax::5} $\upvarphi\equiv\uppsi\to\neg\upvarphi\equiv\neg\uppsi$
\item\label{ax::6} $\upvarphi\equiv\uppsi\to(\upchi\equiv\uptheta\to(\upvarphi\to\upchi)\equiv(\uppsi\to\uptheta))$
\item\label{ax::7}$\upvarphi\equiv\uppsi\to(\upchi\equiv\uptheta\to(\upvarphi\equiv\upchi)\equiv(\uppsi\equiv\uptheta))$
\item\label{ax::8} $\upvarphi\equiv\uppsi\to(\upvarphi\to\uppsi)$. 
\end{enumerate}

\paragraph{Semantics} Let $U\neq\emptyset$, $D\subset U$, and let $\tilde{\neg}: U\longrightarrow U$, $\tilde{\rightarrow} : U \times U \longrightarrow U$, and $\tilde{\equiv} : U \times U \longrightarrow U$ be functions on $U$. An \emph{$\SCI$-model} is a structure $\mathcal{M}=\langle U,D, \tilde{\neg}, \tilde{\rightarrow}, \tilde{\equiv}\rangle$, where $U$ and $D$ are called, respectively, \emph{universe} and \emph{set of designated values}, and the following conditions are satisfied for all $a, b \in U$:
\begin{align}
	\tilde{\neg} a \in D&\qquad\text{iff}\qquad a \notin D\label{cond::model:1}\\
	a \tilde{\rightarrow} b \in D&\qquad\text{iff}\qquad a \notin D\text{ or } b\in D\label{cond::model:2}\\
	a \tilde{\equiv} b\in D&\qquad\text{iff}\qquad a=b.\label{cond::model:3}
\end{align}

\noindent A \emph{valuation} in an $\SCI$- model $\mathcal{M} = \langle U,D, \tilde{\neg}, \tilde{\rightarrow}, \tilde{\equiv}\rangle$ is a function $V:\mathsf{FOR}\longrightarrow U$ such that for all $\upvarphi, \uppsi \in \mathsf{FOR}$ it holds that $V(\neg \upvarphi) = \tilde{\neg} V(\upvarphi)$ and $V(\upvarphi \# \uppsi) = V(\upvarphi) \tilde{\#} V(\uppsi)$, for $\# \in \{\rightarrow, \equiv\}$.
An element $a\in U$ such that $a = V(\upvarphi)$ is called the \emph{denotation of $\upvarphi$}. Interestingly, $\SCI$-model can be defined alternatively as a triple $\mathcal{M} = \langle U,D,V \rangle$, where a valuation $V:{\sf FOR} \longrightarrow U$ needs to satisfy the conditions analogous to (\ref{cond::model:1})--(\ref{cond::model:3}) (for instance, $V(\neg\upvarphi)\in D$ iff $V(\upvarphi)\notin D$ etc.). In the original approach $V$ may as well be defined only for atomic formulas and then lifted up homomorphically to the set of all formulas, like in classical propositional logic. In the latter setting it is not the case, as a valuation defined solely for atoms does usually not have a unique extension to all formulas. We say that a formula $\upvarphi$ is \emph{satisfied} in an $\SCI$-model $\mathcal{M} = \langle U,D, \tilde{\neg}, \tilde{\rightarrow}, \tilde{\equiv}\rangle$ and a valuation $V$ in $\mathcal{M}$, and refer to it as $\mathcal{M}, V \models_{\sf SCI}\upvarphi$, if its denotation belongs to $D$. We call a formula $\upvarphi$ \emph{satisfiable} if it is satisfied in some  $\SCI$-model by some valuation. We say that a formula $\upvarphi$ is \emph{true} in a model $\mathcal{M} = \langle U,D, \tilde{\neg}, \tilde{\rightarrow}, \tilde{\equiv}\rangle$, and refer to it as $\mathcal{M}\models_{\sf SCI}\upvarphi$,  whenever it is satisfied in $\mathcal{M}$ by all the valuations in $\mathcal{M}$.  We call a formula $\upvarphi$ \emph{valid}, and refer to it as $\models_{\sf SCI}\upvarphi$, if it is true in all $\SCI$-models. Note that over the class of models where $D$ and $U\setminus D$ are singletons $\SCI$ collapses to classical propositional logic. In fact all formulas which are $\SCI$-instances of formulas valid in classical  propositional are also valid in $\SCI$. It suffices, however, to take a three-element model to tell $\leftrightarrow$ and $\equiv$ apart, as shown in the following example.
\begin{example}
Although the formula $\neg\neg p \leftrightarrow p$ is a tautology of classical propositional logic, the formula $\neg\neg p \equiv p$ is not valid in $\SCI$. Indeed, consider an $\SCI$-model $\mathcal{M}=\langle U, D, \tilde{\neg}, \tilde{to}, \tilde{\equiv}\rangle$, where $U = \{0, 1, 2\}$, $D = \{1, 2\}$, and the operations $\tilde{\neg}$, $\tilde{\to}$, $\tilde{\equiv}$ are defined by:

\noindent $\tilde{\neg}a = \begin{cases}
0,&\text{if } a\neq 0,\\
1,&\text{otherwise}.
\end{cases}$\ \
$a \tilde{\to} b = \begin{cases}
0,& \begin{minipage}[t]{2cm}if $a \neq 2$ and $b = 0$,\end{minipage}\\
2,& \text{if } a = b,\\
1,& \text{otherwise.}
\end{cases}$\ \
$a \tilde{\equiv} b = \begin{cases}
0,& \text{if } a \neq b\\
a,& \begin{minipage}[t]{2cm}if $a = b$ and $a \neq 0$,\end{minipage}\\
1,& \text{otherwise.}
\end{cases}$

It is easy to verify that such a structure is an SCI-model. Then, the following hold: 
\begin{itemize}	
\item $\tilde{\neg}\tilde{\neg}2 = 1$, and so, $\mathcal{M}$ and a valuation $V$ in $\mathcal{M}$ such that $V(p) = 2$ falsify the formula $\neg\neg p \equiv p$,
\item $1 \tilde{\to} 2 = 1$, but $\tilde{\neg}2 \tilde{\to} \tilde{\neg}1 = 2$, and so, the formula $(p \to q) \equiv (\neg q \to \neg p)$ is not true in $\mathcal{M}$.
\end{itemize}
\end{example}
What is also characteristic of $\SCI$ is that identical formulas can be interchanged within other formulas with not only truth preservation, but also identity preservation. For instance, if $p \equiv (p \to q)$, then $p \equiv ((p \to q)\to q)$, $p \equiv (((p \to q)\to q)\to q)$ and so on. On the other hand, identity of two formulas does not automatically yield identity of their subformulas. For example, if $\neg p \equiv \neg q$, it does not necessarily mean that $p \equiv q$.
It is worth noting that in $\SCI$ we lack the usual equivalence between treating $\land$, $\lor$, and $\leftrightarrow$ as abbreviations involving $\neg$ and $\to$ and treating them as independent connectives whose mutual relations are established axiomatically. For instance, when $\neg(\upvarphi\to\neg\uppsi)$ is just a notational variant for $\upvarphi\land\uppsi$, then $(\upvarphi\land\uppsi)\equiv\neg(\upvarphi\to\neg\uppsi)$ is, of course, ${\sf SCI}$-valid; however, it would not be the case if we regarded $\land$ as a separate connective. Nevertheless, extending our results to other connectives introduced as independent logical constants is a matter of routine.

\section{Tableaux}\label{sect::tableaux}

In this section, we provide a characterization of a sound, complete and terminating \emph{labelled} tableau system for the logic $\SCI$, which we call $\TSCI$. 

Let $\sf L^+$, $\sf L^-$ be countably infinite disjoint sets and let $\mathsf{L}=\mathsf{L}^+\cup\mathsf{L}^-$. We will call an expression $w:\upvarphi$ a \emph{labelled formula}, where $w\in\mathsf{L}$ and $\upvarphi\in\mathsf{FOR}$, and $w$ will be called a \emph{label}. We will abbreviate the set of all labelled formulas by $\sf LF$. Any labels superscribed with `$+$' are restricted to belong to $\mathsf{L}^+$ and labels superscribed with `$-$' to belong to $\mathsf{L}^-$. Labels without a superscript are not restricted. Intuitively, $w$ stands for the denotation of $\upvarphi$ in an intended model. Labels with `$+$' in the superscript denote elements of $D$, whereas labels with superscribed `$-$' represent elements of $U\setminus D$. Thus, expressions of the form $w=v$ or $w\neq v$ reflect, respectively, the equality or distinctness of two denotations. By $\sf Id^+$, $\sf Id^-$ we denote the sets of, respectively, all equalities and all inequalities of labels. Finally, we let $\sf Id=Id^+\cup Id^-$.

A \emph{tableau} generated by the system for the logic $\SCI$ is a \emph{derivation tree} whose nodes are assigned labelled formulas and (in)equality expressions. A simple path $\mathcal{B}$ from the root to a leaf in a tableau $\mathcal{T}$ is called \emph{branch of $\mathcal{T}$}. We will identify a branch $\mathcal{B}$ with the set of labelled formulas and (in)equalities occurring on $\mathcal{B}$.

The rules of our tableau system have the following general form: $\frac{\Upphi}{\Uppsi_1 | \ldots | \Uppsi_n}$, where $\Upphi$ is the set of \emph{premises} and each $\Uppsi_i$, for $i\in\{1,\ldots,n\}$, is a set of \emph{conclusions}. Intuitively, the `$\mid$' symbol should be read as a meta-disjunction. A rule with only one set of conclusions is called a \emph{non-branching} rule. A rule with several sets of conclusions is a \emph{branching rule}. In $\TSCI$ all rules where $\Uppsi_i$, for $i\in\{1,\ldots,n\}$ contain labelled formulas are called \emph{decomposition rules}. All rules with a single equality statement as the conclusion are called \emph{equality rules}. The remaining rules, in which $\bot$ occurs as the conclusion, are referred to as \emph{closure rules}. If we have a decomposition rule $(\sf R)$ with $w:\upvarphi$ as its premise, then $(\sf R)$ is \emph{applicable} to $w:\upvarphi$ occurring on a branch $\mathcal{B}$ if it has not been applied to $w:\upvarphi$ on $\mathcal{B}$ before. Otherwise $w:\upvarphi$ is called \emph{$(\sf R)$-expanded} on $\mathcal{B}$. For an equality rule $(\sf R)$ with $\Upphi$ as the set of premises and $w=v$ as the conclusion, $(\sf R)$ is applicable to $\Upphi\subseteq\mathcal{B}$ if $w=v$ is not present on $\mathcal{B}$. Otherwise $\Upphi$ is $(\sf R)$-expanded on $\mathcal{B}$.
Intuitively, if a set of premises $\Upphi$ is ($\sf R$)-expanded on $\mathcal{B}$, then applying ($\sf R$) to $\Upphi$ would not add any new information to $\mathcal{B}$.

A branch $\mathcal{B}$ of a tableau $\mathcal{T}$ is extended by applying rules of the system to sets of labelled formulas and (in)equality statements that are already on $\mathcal{B}$. A label $w$ is \emph{present} on $\mathcal{B}$ if there exists a formula $\upvarphi$ such that $w:\upvarphi$ occurs on $\mathcal{B}$. Otherwise $w$ is \emph{fresh} on $\mathcal{B}$. A branch $\mathcal{B}$ is called \emph{closed} if one of the closure rules has been applied to it, that is, when an inconsistency occurs on $\mathcal{B}$. A branch that is not closed, is \emph{open}. A branch $\mathcal{B}$ is \emph{fully expanded} if it is closed or no rules are applicable on it. A tableau $\mathcal{T}$ is called \emph{closed} if all of its branches are closed. Otherwise $\mathcal{T}$ is called \emph{open}. We call $\mathcal{T}$ fully expanded if all of its branches are fully expanded.

Analytic tableaux are satisfiability checkers, so a \emph{tableau proof} of a formula $\upvarphi$ is a closed tableau with a labelled formula $w^-:\upvarphi$ at its root. A formula is \emph{tableau-valid} if all tableaux with $w^-:\upvarphi$ at the root are closed. On the other hand, a formula $\varphi$ is \emph{tableau-satisfiable} if there exists an open and fully expanded tableau with a labelled formula $w^+:\upvarphi$ at its root. Note that our notion of tableau-satisfiability matches the usual notion of satisfiability as a failure of finding a proof. Indeed, if a formula $\upvarphi$ is not tableau-valid, that is, there exists a tableau with $w^-:\upvarphi$ at the root which has an open branch, then $\neg\upvarphi$ is tableau-satisfiable. Thus, the standard duality between validity and satisfiability is reflected in the concepts of tableau-validity and tableau-satisfiability.

\begin{figure}[tb!]
	\bgroup\centering	
	($\neg^+$)\quad $\dfrac{w^+:\neg\upvarphi}{v^-:\upvarphi}$\qquad($\neg^-$)\quad $\dfrac{w^-:\neg\upvarphi}{v^+:\upvarphi}$\\[2em]
	($\to^+$)\quad $\dfrac{w^+:\upvarphi\to\uppsi}{\parbox{1.2cm}{\centering $v^-:\upvarphi$\\$u^-:\uppsi$}\,\left|\,\parbox{1.2cm}{\centering $v^-:\upvarphi$\\$u^+:\uppsi$}\right.\,\left|\,\parbox{1.2cm}{\centering $v^+:\upvarphi$\\$u^+:\uppsi$}\right.}$\qquad($\to^-$)\quad$\dfrac{w^-:\upvarphi\to\uppsi}{\parbox{2cm}{\centering $v^+:\upvarphi$\\$u^-:\uppsi$}}$\\[2em]
	($\equiv^+$)\quad $\dfrac{w^+:\upvarphi\equiv\uppsi}{\parbox{1.3cm}{\centering $v^+:\upvarphi$\\$u^+:\uppsi$\\$v^+=u^+$}\,\left|\,\parbox{1.3cm}{\centering $v^-:\upvarphi$\\$u^-:\uppsi$\\$v^-=u^-$}\right.}$\qquad
	($\equiv^-$)\quad$\dfrac{w^-:\upvarphi\equiv\uppsi}{\left.\parbox{1.3cm}{\centering $v^+:\upvarphi$\\$u^+:\uppsi$\\$v^+\neq u^+$}\,\right|\, \parbox{1.2cm}{\centering $v^+:\upvarphi$\\$u^-:\uppsi$}\,\left|\, \parbox{1.2cm}{\centering $v^-:\upvarphi$\\$u^+:\uppsi$}\right.\,\left|\,\parbox{1.3cm}{\centering $v^-:\upvarphi$\\$u^-:\uppsi$\\$v^-\neq u^-$}\right.}$\\[2em]
	($\equiv^{\neg}$)\quad$\dfrac{\parbox{1.2cm}{\centering $\upvarphi\approx\uppsi$\\ $u:\neg\upvarphi$\\$y:\neg\uppsi$}}{u=y}$\qquad
	($\equiv^{\to}$)\quad$\dfrac{\parbox{1.5cm}{\centering $\upvarphi\approx\uppsi$\\ $\upchi\approx\uptheta$\\ $x:\upvarphi\to\upchi$\\$z:\uppsi\to\uptheta$}}{\parbox{1.5cm}{\centering $x=z$}}$\qquad
	($\equiv^{\equiv}$)\quad$\dfrac{\parbox{1.5cm}{\centering $\upvarphi\approx\uppsi$\\ $\upchi\approx\uptheta$\\ $x:\upvarphi\equiv\upchi$\\$z:\uppsi\equiv\uptheta$}}{\parbox{1cm}{\centering $x=z$}}$\qquad
	($\sf F$)\quad$\dfrac{\parbox{1cm}{\centering $w:\upvarphi$\\ $v:\upvarphi$}}{w=v}$\\[2em]
	($\mathsf{sym}$)\quad$\dfrac{w=v}{v=w}$\qquad($\mathsf{tran}$)\quad$\dfrac{\parbox{1.1cm}{\centering$w=v$\\$v=u$}}{w=u}$\qquad($\bot_1$)\quad $\dfrac{\parbox{1cm}{\centering $w=v$\\ $w\neq v$}}{\bot}$\qquad($\bot_2$)\quad $\dfrac{w^+=v^-}{\bot}$\\\egroup
	
	\bigskip
	
	\textsuperscript{1} Labels occurring in conclusions of the rules: ($\neg^+$), ($\neg^-$), ($\to^+$), ($\to^-$), ($\equiv^+$), ($\equiv^-$) are fresh on the branch.\\
	\textsuperscript{2} The abbreviation $\upvarphi\approx\uppsi$ represents the set of three preconditions: $w:\upvarphi$, $v:\uppsi$, $w=v$, for some $w,v\in\mathsf{L}$. Similarly for $\upchi\approx\uptheta$.
	\caption{Tableau system $\TSCI$}
	\label{fig::tableau_calculus}
\end{figure}

\subsection{Tableau system for $\SCI$}\label{sect::tableau_calculus}

The rules presented in Figure~\ref{fig::tableau_calculus} constitute the tableau system $\TSCI$ for the logic $\SCI$. The decomposition rules $(\neg^+)$, $(\neg^-)$, $(\to^+)$, $(\to^-)$, $(\equiv^+)$, $(\equiv^-)$ reflect the semantics of $\neg$, $\to$ and $\equiv$ defined in the conditions~\ref{cond::model:1}--\ref{cond::model:3} from Section~\ref{sect::SCI}. Note that an application of any of these rules introduces to a branch fresh labels for each of the subformulas into which the premise formula is decomposed. By that means, all occurrences of subformulas of the input formula $\upvarphi$ are assigned their unique labels. A few words of extra commentary on the rule $(\equiv^-)$ are in order. It decomposes a formula involving the $\equiv$ connective, which is assumed to be false. By the semantics of $\equiv$ we know that the constituents of the initial $\equiv$-formula have distinct denotations. If these denotations have different polarities, representing different truth values (disjuncts 2 and 3 in the denominator of the rule), then no additional information has to be stored about the distinctness of these denotations. If, on the other hand, the denotations have the same polarity, representing the same truth value (disjuncts 1 and 4 in the denominator of the rule), then extra information is added, namely that the denotations of both formulas are distinct. The rules $(\equiv^\neg)$, $(\equiv^\to)$ and $(\equiv^\equiv)$ are tableau-counterparts of the axioms~\ref{ax::5}, \ref{ax::6}, and \ref{ax::7}, respectively. The rule $(\mathsf{F})$ ensures that a valuation that can be read off from an open branch is a function, i.e., that all denotations assigned to the same formula on a branch are equal. The rules $(\sf sym)$ and $(\sf tran)$ guarantee that equalities appearing on a branch preserve all properties of the $=$-relation. Note that an application of a closure rule to a branch is always a result of transformations of equality statements.
While executing $\TSCI$ we always apply closure rules eagerly, that is, whenever a closure rule can be applied, it should be applied. An example of a tableau proof generated by $\TSCI$ can be found in Figure~\ref{fig::tableau_proof}.
\setlength{\intextsep}{0pt}
\begin{wrapfigure}{r}{.5\linewidth}
	\centering
	\begin{tikzpicture}
	[
	grow                    = down,
	sibling distance        = 6em,
	level distance          = 4em,
	edge from parent/.style = {draw},
	]
	\node [root] {$w^-:\upvarphi\equiv\uppsi\to(\upvarphi\to\uppsi)$}
	child { node [treenode] {$v^+:\upvarphi\equiv\uppsi$\\$u^-:\upvarphi\to\uppsi$}
		child { node [treenode] {$x^+:\upvarphi$\\$y^-:\uppsi$}
			child { node [treenode] {$z^+:\upvarphi$\\$t^+:\uppsi$\\$z^+=t^+$}
				child { node [treenode] {$y^-=t^+$}
					child { node [treenode] {$\bot$}
						edge from parent node [left] {\footnotesize $(\bot_2)$}
					}
					edge from parent node [left] {\footnotesize $(\mathsf{F})$}
				}
				edge from parent node [right=-0.2mm] {\footnotesize $(\equiv^+)$}
			}
			child { node [treenode] {$z^-:\upvarphi$\\$t^-:\uppsi$\\$z^-=t^-$}
				child { node [treenode] {$x^+=z^-$}
					child { node [treenode] {$\bot$}
						edge from parent node [right] {\footnotesize $(\bot_2)$}
					}
					edge from parent node [right] {\footnotesize $(\mathsf{F})$}
				}
			}
			edge from parent node [right] {\footnotesize $(\to^-)$}
		}
		edge from parent node [right] {\footnotesize $(\to^-)$}
	};
	\end{tikzpicture}
	\caption{Tableau proof for the axiom $\upvarphi\equiv\uppsi\to(\upvarphi\to\uppsi)$}
	\label{fig::tableau_proof}
\end{wrapfigure}
The tableau system $\TSCI$ is a user-friendly and elegant solution to the problem most non-labelled systems for $\SCI$ struggle with, namely substitutability of identical formulas within other formulas with identity preservation. In a derivation that can result in yielding conclusions of greater complexity than premises, as shown at the end of Section~\ref{sect::SCI}. It often leads to a loss of subformula property in a deduction system. $\TSCI$, on the other hand, reduces the whole reasoning to a simple equality calculus where only identities or non-identities between labels are substantial for the result of a given derivation. It allows us to circumvent the abovementioned problem by replacing it with a question: are labels representing given formulas equal or distinct?

\subsection{Soundness and completeness}\label{subsect::SoundCompl}

\noindent First, we will prove soundness of the tableau system $\TSCI$.

Let $A, B$ be finite sets such that $A \subseteq \mathsf{LF}$ and $B \subseteq \mathsf{Id}$.  A set  $A \cup B$ is said to be satisfied in an $\SCI$-model $\mathcal{M}= \langle U, D, \tilde{\neg}, \tilde{\rightarrow}, \tilde{\equiv}\rangle$ by a valuation $V$ in $\mathcal{M}$ and a function $f : \mathsf{L} \longrightarrow U$ if and only if the following hold: \begin{enumerate*}[label=(\arabic*)] \item $V(\upvarphi) = f(w)$, for all $w \in \mathsf{L}$ and $\upvarphi \in \mathsf{FOR}$ such that $w : \upvarphi \in A$, \item $f(w) \in D$ iff $w \in \mathsf{L}^+$, for all labels $w$ that occur in $A \cup B$, \item $f(w) = f(v)$, for all $w, v \in \mathsf{L}$ such that $w=v \in B$, \item $f(w) \neq f(v)$, for all $w, v \in \mathsf{L}$ such that $w\neq v \in B$.\end{enumerate*} A set $A \cup B$ is said to be $\SCI$-satisfiable whenever there exist an $\SCI$-model $\mathcal{M}= \langle U, D, \tilde{\neg}, \tilde{\rightarrow}, \tilde{\equiv} \rangle$, a valuation $V$ in $\mathcal{M}$, and a function $f : \mathsf{L} \longrightarrow U$ such that $A \cup B$ is satisfied in $\mathcal{M}$ by $V$ and $f$.   
   
\begin{proposition}\label{labelsat}
For every satisfiable $\SCI$-formula $\upvarphi$ and for all $w^+ \in \mathsf{L}^+$ it holds that $\{w^+ : \upvarphi\}$ is $\SCI$-satisfiable. 
\end{proposition}


\begin{proposition}\label{soundaxiom}
For all  $w, v \in \mathsf{L}$, $w^+ \in \mathsf{L}^+$, and $v^- \in \mathsf{L}^-$, and for all finite $X \subseteq \mathsf{LF} \cup \mathsf{Id}$, the sets $X \cup \{w=v, w \neq v\}$ and  $X \cup \{w^+=v^-\}$ are not $\SCI$-satisfiable.  
\end{proposition}
\noindent Let $(\mathsf{R})$\ $\frac{\Upphi}{\Uppsi_1 | \ldots | \Uppsi_n}$, for $n \geq 1$, be a decomposition or equality rule of the tableau system $\TSCI$. A rule $(\mathsf{R})$ is referred to as \emph{sound} whenever, for every finite set $X \subseteq \mathsf{LF} \cup \mathsf{Id}$, it holds that $X \cup \Upphi$ is $\SCI$-satisfiable iff $X \cup \Upphi \cup \Uppsi_i$ is $\SCI$-satisfiable for some $i \in \{1, \ldots, n\}$. 

\begin{proposition}\label{soundrule}
Decomposition and equality rules of the tableau system $\TSCI$ are sound.
\end{proposition}


\begin{theorem}[Soundness]\label{thm::Soundness}
The tableau system $\TSCI$ is sound, that is, if an $\SCI$ formula $\upvarphi$ is satisfiable, then $\upvarphi$ is tableau-satisfiable.
\end{theorem}

\begin{proof} We prove the contrapositive. Let $\mathcal{T}$ be a closed $\TSCI$-tableau with $w^+ : \upvarphi$ at its root. Then, each branch of $\mathcal{T}$ contains either $w^+ = v^-$ or both $w=v$ and $w \neq v$, for some $w,v \in \mathsf{L}$, $w^+ \in \mathsf{L}^+$, $v^- \in \mathsf{L}^-$. By Proposition~\ref{soundaxiom}, both sets $X\cup \{w^+ = v^-\}$ and $X \cup \{w=v, w\neq v\}$ are not $\SCI$-satisfiable, for any finite set $X \subseteq \mathsf{LF} \cup \mathsf{Id}$. By Proposition~\ref{soundrule}, each application of $\TSCI$-rules preserves $\SCI$-satisfiability. Hence, going from the bottom to the top of the tree $\mathcal{T}$, on each step of the construction of $\TSCI$-tableau we get $\SCI$-unsatisfiable sets. Thus, we can conclude that $w^+ : \upvarphi$ is not $\SCI$-satisfiable, and thus by Proposition~\ref{labelsat} we obtain that $\upvarphi$ is not SCI-satisfiable. Therefore, each satisfiable ${\sf SCI}$-formula $\upvarphi$ is tableau-satisfiable.\end{proof}
  
To prove completeness of the system $\TSCI$ we need to show that if, for a given formula $\upvarphi$, $\TSCI$ does not yield a tableau proof, then $\upvarphi$ is not valid, i.e., there exists a countermodel $\mathcal{M}=\langle U,D,V\rangle$ such that $\mathcal{M}\not\models\upvarphi$.

Suppose that we want to obtain a tableau-proof for a formula $\upvarphi$. To that end, we run the $\TSCI$-tableau algorithm with a labelled formula ${\mathbf{w}^-}: \upvarphi$ at the root of the tableau, for $\mathbf{w}^- \in \mathsf{L}^-$. Suppose that it yields an open tableau as a result. It means that the tableau contains an open and fully expanded branch $\mathcal{B}$. We will demonstrate how to construct a structure $\mathcal{M}_{\mathcal{B}}=\langle U,D, \tilde{\neg}, \tilde{\rightarrow}, \tilde{\equiv}\rangle$ using information stored on $\mathcal{B}$ and show that it actually is an $\SCI$-countermodel falsifying $\upvarphi$. Let $\mathsf{L}_{\mathcal{B}}^+$ be the set of all labels superscribed with `$+$' occurring on $\mathcal{B}$, let $\mathsf{L}_{\mathcal{B}}^-$ be the set of all labels superscribed with `$-$' occurring on $\mathcal{B}$ and let $\mathsf{L}_{\mathcal{B}}=\mathsf{L}_{\mathcal{B}}^+\cup\mathsf{L}_{\mathcal{B}}^-$. Moreover, let $\mathsf{FOR}_{\mathcal{B}}$ be the set of all $\SCI$-formulas $\upvarphi$ such that $w: \upvarphi$ occurs on $\mathcal{B}$, for some $w \in \mathsf{L}_{\mathcal{B}}$. Note that all elements of $\mathsf{FOR}_{\mathcal{B}}$ are subformulas of $\upvarphi$. Before we characterize the construction of $\mathcal{M}_{\mathcal{B}}$, we define a binary relation $\sim\subseteq\mathsf{L}_{\mathcal{B}}\times\mathsf{L}_{\mathcal{B}}$ in the following way:
\begin{center}
	$w\sim v$\qquad iff\qquad\begin{minipage}[m]{0.5\textwidth} $w=v$ occurs on $\mathcal{B}$.
	\end{minipage}
\end{center}

\begin{proposition}\label{prop::equivalence} The relation $\sim$ is an equivalence relation and $(\mathsf{L}_{\mathcal{B}}^+\times\mathsf{L}_{\mathcal{B}}^-)\cap{\sim}=\emptyset$. 
\end{proposition}


Let $\mathsf{ML}_\mathcal{B}^+$ be a set resulting from choosing exactly one label from each element of $({\mathsf{L}_{\mathcal{B}}^+})_{/\sim}$. Sets $\mathsf{ML}_\mathcal{B}^-$ and $\mathsf{ML}_\mathcal{B}$ are defined analogically with the assumption that $\mathbf{w}^- \in \mathsf{ML}_\mathcal{B}^-$, where $\mathbf{w}^-$ is such that $\mathbf{w}^- : \upvarphi$ is at the root of an open tableau. Of course, neither of these sets is uniquely determined. 

\begin{proposition}\label{allsets}
For all $\uppsi \in \mathsf{FOR}$ and $w, v \in \mathsf{L}_{\mathcal{B}}$ the following holds:
\begin{center}
if both $w: \uppsi$ and $v: \uppsi$ belong to $\mathcal{B}$, then $w \sim v$.
\end{center}
\end{proposition}


We say that $w \in \mathsf{ML}_\mathcal{B}$ is \emph{$(\neg)$-closed} whenever there are  $\uppsi \in \mathsf{FOR}$, $u \in \mathsf{ML}_\mathcal{B}$, and $v, t \in \mathsf{L}_{\mathcal{B}}$ such that $w \sim v$, $u \sim t$ and labelled formulas $v: \uppsi$, $t : \neg \uppsi$ belong to $\mathcal{B}$. Let $w, v \in \mathsf{ML}_\mathcal{B}$ and $\# \in \{\rightarrow, \equiv\}$. The pair $(w, v)$ is said to be \emph{$(\#)$-closed} whenever there exist $\uppsi, \uptheta \in \mathsf{FOR}$, $u \in \mathsf{ML}_\mathcal{B}$, and $t, x, y  \in \mathsf{L}_{\mathcal{B}}$ such that $w \sim t$, $v \sim x$, $u \sim y$ and labelled formulas $t : \uppsi$, $x : \uptheta$, $y : (\uppsi \# \uptheta)$ occur on the branch~$\mathcal{B}$.
\medskip

\noindent The \emph{branch structure} $\mathcal{M}_{\mathcal{B}} = \langle U, D, \tilde{\neg}, \tilde{\rightarrow}, \tilde{\equiv}\rangle$ is defined as follows:
	\begin{itemize}
		\item $D=\{w^+\mid w^+\in\mathsf{ML}_\mathcal{B}^+\}\cup \{\mathbf{w}^+\}$, where $\mathbf{w}^+\notin\mathsf{L}_{\mathcal{B}}$
		\item $U=D\cup \mathsf{ML}_\mathcal{B}^-$.
	\end{itemize}
It follows from the above that $U\setminus D=\mathsf{ML}_\mathcal{B}^-$. The operations $\tilde{\neg}, \tilde{\rightarrow}, \tilde{\equiv}$ are defined for all $w, v \in U$ in the following way:	\[\tilde{\neg} w \stackrel{\mathrm{df}}{=} \left\{ \begin{array}{ll} u \in U, &\quad \begin{minipage}[t]{9.2cm} if there are  $\uppsi \in \mathsf{FOR}$ and $v, t \in \mathsf{L}_{\mathcal{B}}$ such that $w = v$, $u = t$, $v: \uppsi$, and $t : \neg \uppsi$ are on $\mathcal{B}$ \end{minipage} \\ \mathbf{w}^+, &\quad \begin{minipage}[t]{7.9cm} if $w$ is not $(\neg)$-closed and $w \not \in  D$ \end{minipage} \\ \mathbf{w}^-, &\quad \mbox{otherwise} \end{array}\right.\] \[w\tilde{\rightarrow} v \stackrel{\mathrm{df}}{=} \left\{ \begin{array}{ll}u \in U, &\quad   \begin{minipage}[t]{8.9cm} if there are $\uppsi, \uptheta \in \mathsf{FOR}$ and $t, x, y  \in \mathsf{L}_{\mathcal{B}}$ such that $w = t$, $v = x$, $u = y$, $t : \uppsi$, $x : \uptheta$, and $y : (\uppsi \rightarrow \uptheta)$ are on $\mathcal{B}$\end{minipage} \\ \mathbf{w}^+, &\quad \begin{minipage}[t]{8.9cm} if $v= \mathbf{w}^+$ or both ($w=\mathbf{w}^+$ and $v \in D$), or it holds that  $(w, v)$ is not $(\rightarrow)$-closed and either $w \not \in D$ or $v \in D$ \end{minipage}\\ \mathbf{w}^-, &\quad \mbox{ otherwise  }    \end{array}\right.\]
\[w\tilde{\equiv} v \stackrel{\mathrm{df}}{=} \left\{ \begin{array}{ll}u \in U, &\quad   \begin{minipage}[t]{8.9cm} if there are $\uppsi, \uptheta \in \mathsf{FOR}$ and $t, x, y  \in \mathsf{L}_{\mathcal{B}}$ such that $w = t$, $v = x$, $u = y$, $t : \uppsi$, $x : \uptheta$, and $y : (\uppsi \equiv \uptheta)$ are on $\mathcal{B}$ \end{minipage} \\ \mathbf{w}^+, &\quad \begin{minipage}[t]{8.9cm} if $w=v$ and either $w = \mathbf{w}^+$ or the pair $(w, v)$ is not $(\equiv)$-closed \end{minipage}\\ \mathbf{w}^-, &\quad \mbox{ otherwise  }     \end{array}\right.\]

\noindent Due to the properties of the sets $\mathsf{ML}_\mathcal{B}^+$ and $\mathsf{ML}_\mathcal{B}^-$, we obtain:

\begin{proposition}\label{universe} The sets $D$ and $U \setminus D$ are non-empty and $D \cap (U \setminus D) = \emptyset$.
\end{proposition}


The following series of results ensure that the operations $\tilde{\neg}$, $\tilde{\to}$, and $\tilde{\equiv}$ reflect the semantics of $\SCI$.

\begin{proposition}\label{negation} $\tilde{\neg}$ is a function on $U$ and for all $w \in U$: 
\begin{enumerate}
\item[$(*)$] $\tilde{\neg}w \in  D$ iff $w \not \in D$.
\end{enumerate}
\end{proposition}


\begin{proposition}\label{implication}
$\tilde{\rightarrow}$ is a function on $U$ and for all $w, v \in U$, the following holds: 
\begin{enumerate}
	\item[$(*)$] $w \tilde{\rightarrow}v \in  D$ iff $w \not \in D$ or $v \in D$.
\end{enumerate}
\end{proposition}


\begin{proposition}\label{equiv}
$\tilde{\equiv}$ is a function on $U$ and for all $w, v \in U$ the following holds: 
\begin{enumerate}
	\item[$(*)$] $w \tilde{\equiv}v \in  D$ iff $w =v$.
\end{enumerate} 
\end{proposition}


\noindent Propositions~\ref{universe}--\ref{equiv} imply:

\begin{proposition}\label{branchmodel}
The structure $\mathcal{M_B}$ is an $\SCI$-model.
\end{proposition}

\noindent In what follows, the structure $\mathcal{M}_{\mathcal{B}}$ will be referred to as \emph{branch model}. 
\medskip

Now, let $V : \mathsf{FOR} \longrightarrow U$ be a function such that for all $p \in \mathsf{AF}$:\label{valuationinB}
\begin{center}
$V(p) = \begin{cases}u \in \mathsf{ML}_\mathcal{B}, &\text{if there is }w \in \mathsf{L}_{\mathcal{B}}\text{ such that }w : p \in \mathcal{B}\text{ and }w \sim u\\
\mathbf{w}^+, &\text{ otherwise }\end{cases}$
\end{center}
and for all $\uppsi, \uptheta \in \mathsf{FOR}$ the following hold:

\medskip
$V(\neg \uppsi) = \tilde{\neg} V(\uppsi)$ 

$V(\uppsi \# \uptheta) = V(\uppsi) \tilde{\#} V(\uptheta)$, for $\# \in \{\rightarrow, \equiv\}$. 

\begin{proposition}\label{valuation} The function $V$ is well defined and it is a valuation in $\mathcal{M}_{\mathcal{B}}$. \end{proposition}


\begin{proposition}\label{formulavalue}
For all $\uppsi \in \mathsf{FOR}$ and $w \in \mathsf{L}_{\mathcal{B}}$ it holds that:  
\begin{enumerate}
	\item[$(*)$] If $w: \uppsi \in \mathcal{B}$, then $w \sim V(\uppsi)$.
\end{enumerate}
\end{proposition}


\begin{theorem}[Completeness]\label{thm::Completeness}
The tableau system $\TSCI$ is complete, that is, if a formula $\upvarphi$ is $\SCI$-valid, then $\upvarphi$ has a tableau proof.
\end{theorem}

\begin{proof} Let $\upvarphi$ be a valid $\SCI$-formula. Suppose that $\upvarphi$ does not have a tableau proof. Then, each $\TSCI$-tableau with $\mathbf{w}^- : \upvarphi$ at its root is open. Let $\mathcal{B}$ be an open and fully expanded branch of an open tableau for $\mathbf{w}^- : \upvarphi$. By Proposition~\ref{branchmodel}, the structure $\mathcal{M_B} = \langle U, D, \tilde{\neg}, \tilde{\rightarrow}, \tilde{\equiv}\rangle$ is an $\SCI$-model. Let $V$ be a valuation in $\mathcal{M_B}$ defined as before Proposition~\ref{valuation} Then, by Proposition~\ref{formulavalue}, $\mathbf{w}^- \sim V(\upvarphi)$, and hence $V(\upvarphi) \not \in D$. Thus, $\upvarphi$ is not true in $\mathcal{M_B}$, which contradicts the assumption that $\upvarphi$ is $\SCI$-valid.\end{proof}

\subsection{Termination}\label{subsect::Termination}

It turns out that the system presented in Section~\ref{sect::tableau_calculus} terminates without any external blocking mechanisms involved which would impose some additional restrictions on rule-application. The only caveat that has to be added to the system is the one that we have already expressed, namely that no rule (\textsf{R}) can be applied to the set of premises that is (\textsf{R})-expanded.

\begin{theorem}\label{thm::termination}
	The tableau system $\TSCI$ is terminating.
\end{theorem}

\begin{proof}
	The argument hinges on two observations. First, the decomposition rules are the only rules that introduce fresh labels to a branch $\mathcal{B}$ of a $\TSCI$-tableau $\mathcal{T}$, and, as mentioned before, on a branch $\mathcal{B}$ each occurrence of a subformula of the initial formula $\upvarphi$ is assigned its unique label. Thus, since an application of any of the above rules decreases the complexity of the processed formula and the rule cannot be applied twice to the same premise, the total number of labels occurring on a branch does not exceed the size of $\upvarphi$ measured as the number of all occurrences of subformulas of $\upvarphi$ (henceforth denoted by $|\upvarphi|$). Secondly, the equality rules can only add equalities between labels to a branch, provided that such an equality statement is not already present thereon. The maximal number of such equalities is quadriatic in the total number of labels occurring on a branch. Thus, for each $\SCI$-formula $\upvarphi$, on any branch $\mathcal{B}$ of a $\TSCI$-tableau for $\upvarphi$, rules are applied at most $|\upvarphi|+|\upvarphi|^2+1$ times, where `$1$' in the formula represents an application of a closure rule. This makes the whole derivation finite.
\end{proof}

\begin{corollary}\label{corollary::polynomialSize}
	For each $\SCI$-formula $\upvarphi$ every branch $\mathcal{B}$ of a $\TSCI$-tableau derivation for $\upvarphi$ is of polynomial size with respect to the size of $\upvarphi$.
\end{corollary}
\noindent Since $\SCI$ contains classical propositional logic, it inherits the \textsc{NP}-lower bound for the satisfiability problem therefrom. Together with membership of $\SCI$-sa\-tis\-fia\-bi\-li\-ty in \textsc{NP} it gives the following:

\begin{theorem}\label{theorem::runningInNP}
	$\TSCI$ is a complexity-optimal decision procedure for the \textsc{NP}-com\-plete problem of $\SCI$-satisfiability.
\end{theorem}

\begin{proof}
	Immediate from Corollary~\ref{corollary::polynomialSize} and the fact that each branching rule of $\TSCI$ is finitely branching.
\end{proof}

\subsection{Limiting the number of labels}

To boost the performance of the system $\TSCI$ we propose a refinement thereof. It consists in limiting the number of fresh labels introduced to a tableau by decomposition rules by introducing an additional condition called \emph{urfather blocking}

Given a formula $\upvarphi$ for which we construct a $\TSCI$-tableau $\mathcal{T}$, for each subformula $\uppsi$ of $\upvarphi$, let's call the first occurrence of a labelled formula $w:\uppsi$ on a branch $\mathcal{B}$ of $\mathcal{T}$ the \emph{$\uppsi$-urfather on $\mathcal{B}$}.
	The system $\TSCI+({\sf UB})$ (\emph{tableau system for {\sf SCI} with urfather blocking}) is composed of the rules of $\TSCI$ and an additional constraint:
	\begin{enumerate}[leftmargin=*,labelindent=13pt]
		\item[$(\sf{UB})$] For each labelled formula $w:\upvarphi$ that occurs on a branch $\mathcal{B}$, no decomposition rule can be applied to $w:\upvarphi$ unless it is the $\upvarphi$-urfather on $\mathcal{B}$.
	\end{enumerate}
It turns out that augmenting $\TSCI$ with $\sf (UB)$ does not lead to any unwanted consequences such as giving up the completeness.

\begin{proposition}\label{prop::UrfatherProof}
	For every $\sf SCI$-formula $\upvarphi$, if $\upvarphi$ has a $\sf \TSCI$-tableau proof, then $\upvarphi$ has $\sf TC_{SCI}+(UB)$-tableau proof. 
\end{proposition}


\begin{theorem}
	$\TSCI+({\sf UB})$ is sound, complete, terminating, and complexity-op\-ti\-mal for $\SCI$-satisfiability.
\end{theorem}

\begin{proof}
	The soundness of $\TSCI+({\sf UB})$ straightforwardly follows from the soundness of $\TSCI$ and the fact that both systems share the full set of rules. The argument for termination of $\TSCI+(UB)$ and complexity-optimality of $\TSCI+(UB)$ for $\SCI$-satisfiability goes along the same lines as the proofs of Theorems~\ref{thm::termination} and~\ref{theorem::runningInNP}, and rests on the fact that, for each formula $\upvarphi$, a $\TSCI+({\sf UB})$-tableau contains at most as many labels as a $\TSCI$-tableau. The completeness of $\\TSCI+(UB)$ is a direct consequence of Proposition~\ref{prop::UrfatherProof} and Theorem~\ref{thm::Completeness}.
\end{proof}

\section{Implementation}\label{sect::implementation}

\subsection{Overview}
We have
written proof-of-concept type implementations of
the labelled tableau system described in the present article
and its variant with urfather blocking,
as well as a dual-tableau-based theorem prover for~$\SCI$ based on the system from~\cite{jgp07}.
Since the last system does not enjoy the termination property, the implementation relies on heuristics in this respect. All three provers are implemented in the Haskell language
using similar programming techniques
in a casual manner,
without any serious attempt to optimize the code
or to test it extensively,
as the programs are only intended as temporary aids to ongoing research.

In testing,
the labelled-tableau provers
turned out to need drastically more computing resources
even in many quite modest test cases.
For instance,
the axiom
$((\pv{p} \equiv \pv{q}) \land (\pv{r} \equiv \pv{s})) \to ((\pv{p} \equiv \pv{r}) \equiv (\pv{q} \equiv \pv{s}))$
generates a labelled tableau of depth~37 consisting of 619~nodes,
which urfather blocking reduces to depth~33 and 555~nodes,
while
the tree of the dual-tableau prover has depth~18 and only 67~nodes.
The difference appears to be mostly due to
the large branching factor of the identity rules of the labelled-tableau system.
However, in some test cases the labelled-tableau system
yields a smaller tree than the other prover.
In general,
the labelled tableau method seems to tolerate relatively well
formulas
consisting of a large number of very simple identitities.

\subsection{Technical notes}
Unlike the abstract tree described above,
each node of which contains only a single labelled formula,
each node of the tree built by the program
contains a list of all the labelled formulas encountered so far on the branch.
This allows the program to freely manipulate the list
to keep track of what rules have already been applied to which formulas.
There are three main types of nodes:
normal nodes, identity nodes, and leaves.
First, the decomposition rules
are applied in normal nodes.
Once they have been applied to exhaustion,
the tree is extended with identity nodes,
in which the identity rules are applied.
At any point, one of the closure rules $(\bot_1)$~or~$(\bot_2)$ can be applied
to append a special closure leaf node.
An open leaf node is appended
whenever there are no more rules to apply in an identity node
and the branch remains open. 

\subsection{Test results}
We found a randomly generated provable
$\SCI$-formula that turned out to be somewhat challenging to an earlier prover.
The formula, which we will call the $\upvarphi$ here, looks as follows:
\begin{align*}
	(((\pv{q} \equiv \pv{p}) &\rightarrow (\pv{p} \rightarrow \pv{r}))
	\equiv ((\pv{p} \rightarrow (\pv{p} \leftrightarrow \pv{p})) \equiv \pv{p}))\\
	&\rightarrow (((\pv{r} \land \pv{p}) \leftrightarrow (\pv{p} \equiv \pv{p}))
	\lor ((\pv{p} \land \pv{p}) \lor \lnot \pv{q}))
\end{align*}
We denote by~$\uppsi$
the formula obtained
by replacing each occurrence of~$\pv{p}$ in~$\upvarphi$ by~$\upvarphi$ itself.
We defined a provability-preserving transformation~$T$
that turns an  $\SCI$-formula into a Horn clause
consisting of very simple identities.

We present the results of attempting to prove the formulas
$\upvarphi$, $\lnot \upvarphi$, $\uppsi$, $\lnot \uppsi$, $T(\upvarphi)$,
and~$T(\lnot\upvarphi)$.
These are chosen to illustrate some of the variety of outcomes we observed.
As noted above, $\upvarphi$~is provable,
and therefore also $\uppsi$~and~$T(\upvarphi)$ are provable.
The results are of the form {\it depth/size,}
where {\it depth} is the maximal branch length
and {\it size} is the number of nodes in the entire tree.
There are entries for the dual-tableau-based prover ($\sf DT_{SCI}$),
the current labelled-tableau prover ($\TSCI$),
and the same with the urfather blocking condition ($\TSCI + ({\sf UB})$).
Several entries are missing due to exhaustion of memory
(the programs were tested on a machine with 8GB of RAM;
adding several gigabytes of swap space did not make a difference).
\begin{center}
	\setlength{\tabcolsep}{8pt}
\begin{tabular}{c|rr|rr|rr}
	\textbf{Formula}&\multicolumn{2}{c|}{$\sf DT_{\SCI}$}&\multicolumn{2}{c|}{$\TSCI$}&\multicolumn{2}{c}{$\TSCI + ({\sf UB})$}\\
	\hline
	&\textbf{depth}&\textbf{size}&\textbf{depth}&\textbf{size}&\textbf{depth}&\textbf{size}\\[1ex]
	$\upvarphi$&27&299&37&4724&32&4659\\
	$\neg\upvarphi$&12&42&202&111539&106&95724\\
	$\uppsi$&61&17729&$-$&$-$&46&3023804\\
	$\neg\uppsi$&42&602&$-$&$-$&$-$&$-$\\
	$T(\upvarphi)$&$-$&$-$&143&40230&106&34158\\
	$T(\neg\upvarphi)$&$-$&$-$&529&52789&490&46153
\end{tabular}
\end{center}

\section{Conclusions}\label{sect::Conclusions}

In this paper we introduced the system $\TSCI$ which is the first complexity-optimal decision procedure for the logic $\SCI$ devised in the paradigm of labelled tableaux. $\TSCI$ is conceptually simple and directly reflects the semantics of the logic. The reasoning performed in $\TSCI$ has two components: decomposition and equality reasoning. Interestingly, it is the latter that is responsible for closing tableau branches, and thus, yielding tableau proofs for formulas. In this respect $\TSCI$ is based on similar conceptual foundations as calculi generated by the tableau-synthesis framework from~\cite{ST11}.We provided an implementation of $\TSCI$ and a variant with urfather blocking, and we compared their performance with the performance of another implemented deduction system for $\SCI$ which has not been proven to be terminating or complete. There was no unique winner; the new system was better at dealing with formulas with complex networks of identities, while the old, unproven system handled other types of formulas better. Urfather blocking yielded modest reductions in depth and total size.

In future research we want to address three main problems. First, we would like to optimize our tableau algorithm by introducing further refinements to it, such as decreasing the branching factor of the rule $(\to^+)$ and, by that means, making it ``information-deleting''. Some prelimiary results on the implementation of $\TSCI$ with the modified rule $(\to^+)$ show a promising reduction of the size of generated tableaus. Moreover, we plan to search for heuristics and rule-application strategies which would, too, allow to minimize the size of tableaux yielded by $\TSCI$ for certain classes of formulas. It seems that it is not always necessary to fully decompose the input formula before performing any equality reasoning, if a contradiction is to be reached on a branch. Secondly, we would like to develop the dual-tableau systems from~\cite{jgp07} and \cite{jgpwelle} to full-fledged decision procedures, implement them, and compare the performance of all three algorithms on an extensive set of various $\SCI$-formulas. Thirdly, we intend to extend the labelled tableaux-based approach presented in this paper to other non-Fregean logics, both classical (such as modal non-Fregean logics) and deviant (such as intuitionistic or many-valued non-Fregean logics, or Grzegorczyk's logic). Finally, we would like to take a closer look at various normal forms of $\sf SCI$ formulas, one of which was mentioned in Section~\ref{sect::implementation}, and decide in what cases it pays off to transform a formula into a normal form before running a decision procedure, rather than running it directly on the initial formula.


\newpage

\begin{subappendices}
	\renewcommand{\thesection}{\Alph{section}}
	\section{Omitted proofs}
	
	\subsection{Proof of Proposition~\ref{labelsat}}
	
	\begin{repproposition}{labelsat}
		For every satisfiable $\mathsf{SCI}$-formula $\upvarphi$ and for all $w^+ \in \mathsf{L}^+$ it holds that $\{w^+ : \upvarphi\}$ is $\mathsf{SCI}$-satisfiable. 
	\end{repproposition}
	
	\begin{proof} Let $\upvarphi$ be a satisfiable $\mathsf{SCI}$-formula. Then, there exist an $\mathsf{SCI}$-model $\mathcal{M}= \langle U, D, \tilde{\neg}, \tilde{\rightarrow}, \tilde{\equiv}\rangle$ and a valuation $V$ in $\mathcal{M}$ such that $V(\upvarphi) \in D$. Let $f : \mathsf{L} \longrightarrow U$ be such that $f(w^+) = V(\upvarphi)$. Clearly, $\{w+ : \upvarphi\}$ is satisfied in $\mathcal{M}$ by $V$ and $f$, so it is $\mathsf{SCI}$-satisfiable. \end{proof}
	
	\subsection{Proof of Proposition~\ref{soundaxiom}}
	
	\begin{repproposition}{soundaxiom}
		For all  $w, v \in \mathsf{L}$, $w^+ \in \mathsf{L}^+$, and $v^- \in \mathsf{L}^-$, and for all finite $X \subseteq \mathsf{LF} \cup \mathsf{Id}$, the sets $X \cup \{w=v, w \neq v\}$ and  $X \cup \{w^+=v^-\}$ are not $\mathsf{SCI}$-satisfiable.  
	\end{repproposition}
	
	\begin{proof} Let $w, v \in \mathsf{L}$, $w^+ \in \mathsf{L}^+$, and $v^- \in \mathsf{L}^-$. Let $\mathcal{M}= \langle U, D, \tilde{\neg}, \tilde{\rightarrow}, \tilde{\equiv} \rangle$ be an $\mathsf{SCI}$-model, let $V$ be a valuation in $\mathcal{M}$, and let $f : \mathsf{L} \longrightarrow U$. Then, if $f(w)=f(v)$, then $f(w) \neq f(v)$ does not hold, so it cannot be the case that $f$ satisfies both conditions (3) and (4) of the definition of $\mathsf{SCI}$-satisfiability. Hence, $X \cup \{w=v, w \neq v\}$ is not $\mathsf{SCI}$-satisfiable. Moreover, if $X \cup \{w^+=v^-\}$ is $\mathsf{SCI}$-satisfiable, then by the condition (3), $f(w^+) = f(v^-)$, but then the condition (2) of the definition of $\mathsf{SCI}$-satisfiability does not hold. Therefore, $X \cup \{w^+=v^-\}$ is not $\mathsf{SCI}$-satisfiable.
	\end{proof}

	\subsection{Proof of Proposition~\ref{soundrule}}

	\begin{repproposition}{soundrule}
	Decomposition and equality rules of the tableau calculus $\mathsf{TC}_{\mathsf{SCI}}$ are sound.
	\end{repproposition}

	\begin{proof} By way of example, we will prove the proposition for the rules $(\neg^+)$, $(\rightarrow^-)$, $(\equiv^-)$, $(\equiv^{\equiv})$, and $(\mathsf{F})$. In what follows, we assume that $X \subseteq \mathsf{LF} \cup \mathsf{Id}$ is finite, $\upvarphi, \uppsi, \uptheta, \upchi$ are $\mathsf{SCI}$-formulas, $w^+, v^+, u^+ \in \mathsf{L}^+$, $w^-, v^-, u^- \in \mathsf{L}^-$, and $w, v, u, y, x, z \in \mathsf{L}$.
	
	\medskip
	\noindent \emph{The rule $(\neg^+)$}
	
	\noindent Assume $X \cup \{w^+ : \neg \upvarphi\}$ is $\mathsf{SCI}$-satisfiable, that is, there exist an $\mathsf{SCI}$-model $\mathcal{M}= \langle U, D, \tilde{\neg}, \tilde{\rightarrow}, \tilde{\equiv}\rangle$, a valuation $V$ in $\mathcal{M}$, and a function $f : \mathsf{L} \longrightarrow U$ such that: $V(\neg \upvarphi) = f(w^+)$ and $f(w^+) \in D$. Hence, by the semantics of $\mathsf{SCI}$, $V(\upvarphi) \not \in D$. Let $f' : \mathsf{L} \longrightarrow U$ be such that $f'(v^-) = V(\upvarphi)$ and $f'(w) =f(w)$, for all $w$ that occur in $X \cup \{w^+ : \neg \upvarphi\}$. Since $v^-$ is a fresh label, that is, it does not occur in $X \cup \{w^+ : \neg \upvarphi\}$, the function $f'$ is well defined. Moreover, $V(\upvarphi)=f'(v^-) \not \in D$ and since $v^- : \upvarphi \not \in X$, the set $X \cup \{w^+ : \neg \upvarphi\} \cup \{v^- : \upvarphi\}$ is satisfied in $\mathcal{M}$ by $V$ and $f'$. Thus, the rule $(\neg^+)$ is sound.

	\medskip
	\noindent \emph{The rule $(\rightarrow^-)$}
	
	\noindent Assume $X \cup \{w^- : \upvarphi \rightarrow \uppsi\}$ is $\mathsf{SCI}$-satisfiable. Then, there exist an $\mathsf{SCI}$-model $\mathcal{M}= \langle U, D, \tilde{\neg}, \tilde{\rightarrow}, \tilde{\equiv}\rangle$, a valuation $V$ in $\mathcal{M}$, and a function $f : \mathsf{L} \longrightarrow U$ such that: $V(\upvarphi \rightarrow \uppsi) = f(w^-)$ and $f(w^-) \not \in D$. Hence, by the semantics of $\mathsf{SCI}$, $V(\upvarphi) \in D$ and $V(\uppsi) \not \in D$. Let $f' : \mathsf{L} \longrightarrow U$ be such that $f'(v^+) = V(\upvarphi)$, $f'(u^-) = V(\uppsi)$, and $f'(w) =f(w)$, for all $w$ that occur in $X \cup \{w^- : \upvarphi \rightarrow \uppsi\}$. As $v^+$ and $u^-$ are fresh labels, the function $f'$ is well defined. Then, $V(\upvarphi) = f'(v^+) \in D$ and $V(\uppsi) = f'(u^-) \not \in D$. Therefore, $X \cup \{w^- : \upvarphi \rightarrow \uppsi\} \cup \{v^+ : \upvarphi, u^- : \uppsi\}$ is satisfied in $\mathcal{M}$ by $V$ and $f'$, so the rule $(\rightarrow^-)$ is sound.    
	
	\medskip
	\noindent \emph{The rule $(\equiv^-)$}
	
	\noindent Assume $X \cup \{w^- : \upvarphi \equiv \uppsi\}$ is $\mathsf{SCI}$-satisfiable. Then, there exist an $\mathsf{SCI}$-model $\mathcal{M}= \langle U, D, \tilde{\neg}, \tilde{\rightarrow}, \tilde{\equiv}\rangle$, a valuation $V$ in $\mathcal{M}$, and a function $f : \mathsf{L} \longrightarrow U$ such that $V(\upvarphi \equiv \uppsi) = f(w^-) \not \in D$. Hence, $V(\upvarphi) \neq V(\uppsi)$. Let us assume that $V(\upvarphi), V(\uppsi) \in D$. Then, let $f' : \mathsf{L} \longrightarrow U$ be such that $f'(v^+) = V(\upvarphi)$, $f'(u^+) = V(\uppsi)$, and $f'(w) =f(w)$, for all $w$ that occur in $X \cup \{w^- : \upvarphi \equiv \uppsi\}$. Labels $v^+$ and $u^+$ are fresh, so $f'$ is well defined. Since $V(\upvarphi) \neq V(\uppsi)$, we obtain $f'(v^+) \neq f'(u^+)$. Furthermore, $f'(v^+), f'(u^+) \in D$. Therefore, $X \cup \{w^- : \upvarphi \equiv \uppsi\} \cup \{v^+ : \upvarphi, u^+ : \uppsi, v^+ \neq u^+\}$ is satisfied in $\mathcal{M}$ by $V$ and $f'$. Now, assume that $V(\upvarphi) \in D$ and $V(\uppsi) \not \in D$. Let $f' : \mathsf{L} \longrightarrow U$ be such that $f'(v^+) = V(\upvarphi)$, $f'(u^-) = V(\uppsi)$, and $f'(w) =f(w)$, for all $w$ that occur in $X \cup \{w^- : \upvarphi \equiv \uppsi\}$. Labels $v^+, u^-$ do not occur in $X \cup \{w^- : \upvarphi \equiv \uppsi\}$, so $f'$ is well defined. Then, $f'(v^+) \in D$ and $f'(u^-) \not \in D$. Therefore, $X \cup \{w^- : \upvarphi \equiv \uppsi\} \cup \{v^+ : \upvarphi, u^- : \uppsi\}$ is satisfied in $\mathcal{M}$ by $V$ and $f'$. In a similar way, we can prove that if $V(\upvarphi), V(\uppsi) \not \in D$ (resp. $V(\upvarphi) \not \in D$ and $V(\uppsi) \in D$), then $X \cup \{w^- : \upvarphi \equiv \uppsi\} \cup \{v^- : \upvarphi, u^- : \uppsi, v^- \neq u^-\}$ (resp. $X \cup \{w^- : \upvarphi \equiv \uppsi\} \cup \{v^- : \upvarphi, u^+ : \uppsi\}$) is $\mathsf{SCI}$-satisfiable. Therefore, if the set $X \cup \{w^- : \upvarphi \equiv \uppsi\}$ is $\mathsf{SCI}$-satisfiable, then either $X \cup \{w^- : \upvarphi \equiv \uppsi\} \cup \{v^+ : \upvarphi, u^+ : \uppsi, v^+ \neq u^+\}$ or $X \cup \{w^- : \upvarphi \equiv \uppsi\} \cup \{v^+ : \upvarphi, u^- : \uppsi\}$ or $X \cup \{w^- : \upvarphi \equiv \uppsi\} \cup \{v^- : \upvarphi, u^+ : \uppsi\}$ or $X \cup \{w^- : \upvarphi \equiv \uppsi\} \cup \{v^- : \upvarphi, u^- : \uppsi, v^- \neq u^-\}$ is $\mathsf{SCI}$-satisfiable, from which it follows that the rule $(\equiv^-)$ is sound.

	\medskip
	\noindent \emph{The rule $(\equiv^{\equiv})$}
	
	\noindent Assume that $\Uppi = X \cup \{w: \upvarphi, v: \uppsi, w=v,  u: \upchi, y: \uptheta, u=y, x : \upvarphi \equiv \upchi, z : \uppsi \equiv \uptheta\}$ is satisfied in an $\mathsf{SCI}$-model $\mathcal{M}= \langle U, D, \tilde{\neg}, \tilde{\rightarrow}, \tilde{\equiv}\rangle$ by a valuation $V$ in $\mathcal{M}$ and a function $f : \mathsf{L} \longrightarrow U$. Then, $f(w) = V(\upvarphi)=V(\uppsi) = f(v)$, $f(u) = V(\upchi)=V(\uptheta) = f(y)$, $f(x)=V(\upvarphi\equiv \upchi)$, and $f(z) = V(\uppsi \equiv \uptheta)$. By the semantics of $\mathsf{SCI}$, if $V(\upvarphi)=V(\uppsi)$ and $V(\upchi)=V(\uptheta)$, then $V(\upvarphi \equiv \upchi)=V(\uppsi\equiv\uptheta)$, so we know that $f(x) = f(z)$. Hence, the set $\Uppi \cup \{x=z\}$ is satisfied in $\mathcal{M}$ by $V$ and $f$, so the rule $(\equiv^{\equiv})$ is sound. 
	
	\medskip
	\noindent \emph{The rule $(\mathsf{F})$}
	
	\noindent Assume that $X \cup \{w: \upvarphi, v : \upvarphi\}$ is  satisfied in an $\mathsf{SCI}$-model $\mathcal{M}= (U, D, \tilde{\neg}, \tilde{\rightarrow}, \tilde{\equiv})$ by a valuation $V$ in $\mathcal{M}$ and a function $f : \mathsf{L} \longrightarrow U$. Then, $f(w) = V(\upvarphi)=f(v) = V(\upvarphi)$, so $f(w) = f(v)$. Therefore, $X \cup \{w: \upvarphi, v : \upvarphi, w=v\}$ is  satisfied in $\mathcal{M}$ by $V$ and $f$,  so the rule $(\mathsf{F})$ is sound.\end{proof}

	\subsection{Proof of Proposition~\ref{prop::equivalence}}

	\begin{repproposition}{prop::equivalence} The relation $\sim$ is an equivalence relation and $(\mathsf{L}_{\mathcal{B}}^+\times\mathsf{L}_{\mathcal{B}}^-)\cap{\sim}=\emptyset$. 
	\end{repproposition}
	
	\begin{proof} Let $w \in \mathsf{L}_{\mathcal{B}}$. Then, there exists a formula $\uppsi$ such that $w : \uppsi$ occurs on $\mathcal{B}$. Thus, the rule $(\mathsf{F})$ applies to $w : \uppsi$, $w : \uppsi$, so $w = w$ must be on $\mathcal{B}$, that is $\sim$ is reflexive. Assume $w \sim v$, for some $w, v \in \mathsf{L}_{\mathcal{B}}$. Then, $w=v$ is on $\mathcal{B}$, and thus, by the rule $(\mathsf{sym})$, also $v = w$ belongs to $\mathcal{B}$, so $v \sim w$. Hence, the relation $\sim$ is symmetric. Assume $w \sim v$ and $v \sim u$. Then, $w=v$ and $v=u$ belong to $\mathcal{B}$, and by the rule $(\mathsf{tran})$, $w=u$ is on $\mathcal{B}$, that is $w \sim u$. Therefore, $\sim$ is transitive. Moreover, it cannot be the case that  $w^+ \sim v^-$, for some $w^+ \in \mathsf{L}^+$, $v^- \in \mathsf{L}^-$, since otherwise $w^+ = v^-$ would belong to $\mathcal{B}$ and the branch $\mathcal{B}$ would have to be closed by an application of the rule $(\bot_2)$, which contradicts the assumption about the openness of $\mathcal{B}$. Therefore, $(\mathsf{L}_{\mathcal{B}}^+\times\mathsf{L}_{\mathcal{B}}^-)\cap{\sim}=\emptyset$. \end{proof}
	
	\subsection{Proof of Proposition~\ref{allsets}}
	
	\begin{repproposition}{allsets}
		For all $\uppsi \in \mathsf{FOR}$ and $w, v \in \mathsf{L}_{\mathcal{B}}$ the following holds:
		\begin{center}
			if both $w: \uppsi$ and $v: \uppsi$ belong to $\mathcal{B}$, then $w \sim v$.
		\end{center}
	\end{repproposition}
	
	\begin{proof} Let $\uppsi \in \mathsf{FOR}$ and $w, v \in \mathsf{L}_{\mathcal{B}}$ be such that $w: \uppsi$ and $v: \uppsi$ belong to $\mathcal{B}$. Then, the rule $(\mathsf{F})$ applies to $w: \uppsi$ and $v: \uppsi$, so by the openness and expandedness of $\mathcal{B}$,  $w=v$ must belong to $\mathcal{B}$. Therefore, $w\sim v$.\end{proof}
	
	\subsection{Proof of Proposition~\ref{negation}}
	
	\begin{repproposition}{negation} $\tilde{\neg}$ is a function on $U$ and for all $w \in U$: 
		\begin{enumerate}
			\item[$(*)$] $\tilde{\neg}w \in  D$ iff $w \not \in D$.
		\end{enumerate}
	\end{repproposition}
	
	\begin{proof} Let $w \in U$. By the definition of $\tilde{\neg}$, if $w = \mathbf{w}^+$ or $w$ is not $(\neg)$-closed, then $\tilde{\neg} w$ has exactly one value. If $w$ is $(\neg)$-closed, then there are $\uppsi \in \mathsf{FOR}$, $u \in \mathsf{ML}_\mathcal{B}$, and $v, t \in \mathsf{L}_{\mathcal{B}}$ such that $w =v$, $u = t$, $v: \uppsi$, $t : \neg \uppsi$ belong to $\mathcal{B}$, and thus, by the definition of $\tilde{\neg}$, we have $\tilde{\neg} w = u$. Now, we will show that this value is unique. Suppose there are $u, u' \in \mathsf{ML}_\mathcal{B}$ such that $\tilde{\neg} w = u$ and $\tilde{\neg}w = u'$. Then, by the definition of $\tilde{\neg}$, there are  $\uppsi, {\uppsi}' \in \mathsf{FOR}$, $v, v', t, t' \in \mathsf{L}_{\mathcal{B}}$ such that $w = v$, $w = v'$, $u = t$, $u' = t'$, $v: \uppsi$, $t : \neg \uppsi$ and $v': {\uppsi}'$, $t' : \neg {\uppsi}'$ are on $\mathcal{B}$. Since $w=v$ and $w = v'$, by the rules $(\mathsf{sym})$ and $(\mathsf{tran})$, we obtain that $v=v'$ is on $\mathcal{B}$. Thus, the rule $(\equiv^{\neg})$ applies to $v: \uppsi$, $v': {\uppsi}'$, $v=v'$, $t : \neg \uppsi$, $t' : \neg {\uppsi}'$, which means that $t=t'$ belongs to $\mathcal{B}$, and thus $t \sim t'$. Since $t \sim u$, $t' \sim u'$, and $t \sim t'$, we obtain that $u \sim u'$, which ends the proof of the uniqueness of $\tilde{\neg}$. 
		
	Now, we will show that $\tilde{\neg}$ satisfies $(*)$. It straightforwardly follows from the definition of $\tilde{\neg}$ that $(*)$ holds if $w = \mathbf{w}^+$ or $w$ is not $(\neg)$-closed. So let us assume that $w \in \mathsf{ML}_\mathcal{B}$ is $(\neg)$-closed. Then, there are  $\uppsi \in \mathsf{FOR}$, $u \in \mathsf{ML}_\mathcal{B}$, and $v, t \in \mathsf{L}_{\mathcal{B}}$ such that $w =v$, $u = t$, $v: \uppsi$, $t : \neg \uppsi$ belong to $\mathcal{B}$. Assume $u \in D$, that is, $u \in \mathsf{L}_{\mathcal{B}}^+$. Since $u \sim t$, we have $t \in \mathsf{L}_{\mathcal{B}}^+$, so the rule $(\neg^+)$ applies to $t : \neg \uppsi$. Thus, $v^- : \uppsi$ is on $\mathcal{B}$, for some $v^- \in \mathsf{L}_{\mathcal{B}}^-$. Hence, both $v : \uppsi$ and $v^- : \uppsi$ are on $\mathcal{B}$, and thus, by Proposition~\ref{allsets}, we obtain $v^- \sim v$. Since $w \sim v$, $v \sim v^-$, and $v^- \in \mathsf{L}^{-}$, we obtain $w \in \mathsf{ML}_\mathcal{B}^- = U \setminus D$. Therefore, $w \not \in D$. On the other hand, if $u \not \in D$, then $u \in \mathsf{L}_{\mathcal{B}}^-$ and the rule $(\neg^-)$ applies to $t : \neg \uppsi$. Thus, $v^+ : \uppsi$ is on $\mathcal{B}$, for some $v^+ \in \mathsf{L}_{\mathcal{B}}^+$. Since both $v : \uppsi$ and $v^+ : \uppsi$ are on $\mathcal{B}$, by Proposition~\ref{allsets}, we get $v^+ \sim v$, so $v^+ \sim w$. Hence, $w \in \mathsf{ML}_\mathcal{B}^+ \subseteq D$. Therefore, $w \in D$.\end{proof}

	\subsection{Proof of Proposition~\ref{implication}}

	\begin{repproposition}{implication}
		$\tilde{\rightarrow}$ is a function on $U$ and for all $w, v \in U$, the following holds: 
		\begin{enumerate}
			\item[$(*)$] $w \tilde{\rightarrow}v \in  D$ iff $w \not \in D$ or $v \in D$.
		\end{enumerate}
	\end{repproposition}

	\begin{proof} Let $w, v \in U$. By the definition of $\tilde{\rightarrow}$, if $w = \mathbf{w}^-$ or $v = \mathbf{w}^+$ or $(w, v)$ is not $(\rightarrow)$-closed, then $w \tilde{\rightarrow} v$ has exactly one value. So let $(w, v)$ be $(\rightarrow)$-closed. Then, there are  $\uppsi, \uptheta \in \mathsf{FOR}$, $u \in \mathsf{ML}_\mathcal{B}$, and $t, x, y  \in \mathsf{L}_{\mathcal{B}}$ such that $w \sim t$, $v \sim x$, $u \sim y$ and labelled formulas $t : \uppsi$, $x : \uptheta$, $y : (\uppsi \rightarrow \uptheta)$ occur on the branch $\mathcal{B}$. Then, $w=t$, $v=x$, $u=y$ belong to $\mathcal{B}$, and by the definition of $\tilde{\rightarrow}$, we obtain $(w\tilde{\rightarrow}v) = u$. Now, we will show that the value $w \tilde{\rightarrow} v$ is unique. Suppose there are $u,u' \in \mathsf{ML}_\mathcal{B}$ such that $(w\tilde{\rightarrow}v) = u$ and $(w\tilde{\rightarrow}v) = u'$. Then, there are  $\uppsi, {\uppsi}', \uptheta, {\uptheta}' \in \mathsf{FOR}$ and $t, t', x, x', y, y'  \in \mathsf{L}_{\mathcal{B}}$ such that $w = t$, $w =t'$, $v = x$, $v=x'$, $u = y$, $u' = y'$, $t : \uppsi$, $x : \uptheta$, $y : (\uppsi \rightarrow \uptheta)$, and $t' : {\uppsi}'$, $x' : {\uptheta}'$, $y' : ({\uppsi}' \rightarrow {\uptheta}')$ occur on the branch $\mathcal{B}$. By the rule $(\mathsf{sym})$ and $(\mathsf{tran})$, we obtain that $t=t'$ and $x = x'$ are on $\mathcal{B}$, so the rule $(\equiv^{\rightarrow})$ had to be applied to $t : \uppsi$, $t' : {\uppsi}'$, $t=t'$, $x : \uptheta$, $x' : {\uptheta}'$, $x=x'$, $y : (\uppsi \rightarrow \uptheta)$, $y' : ({\uppsi}' \rightarrow {\uptheta}')$, which means that $y=y'$ is on $\mathcal{B}$, that is, $y \sim y'$. Hence, $u \sim u'$, so  $\tilde{\rightarrow}$ is a function on $U$. 
		
	Now, we will show that $\tilde{\rightarrow}$ satisfies $(*)$. Let $w, v \in U$. Observe that  if $w = \mathbf{w}^+$ or $v = \mathbf{w}^+$ or $(w, v)$ is not $(\rightarrow)$-closed, then $(*)$ follows straightforwardly from the definition of $\tilde{\rightarrow}$. Thus, assume that $(w, v)$ is $(\rightarrow)$-closed. Then, $w, v \in \mathsf{ML}_\mathcal{B}$ and there are  $\uppsi, \uptheta \in \mathsf{FOR}$, $u \in \mathsf{ML}_\mathcal{B}$, and $t, x, y  \in \mathsf{L}_{\mathcal{B}}$ such that $w = t$, $v = x$, $u = y$, $t : \uppsi$, $x : \uptheta$, $y : (\uppsi \rightarrow \uptheta)$ occur on the branch $\mathcal{B}$. Let $(w \tilde{\rightarrow} v) = u \in D$. Then, $u \in \mathsf{ML}_\mathcal{B}^+$, that is $y \in \mathsf{L}^+$. So the rule $(\rightarrow^+)$ applies to $y : (\uppsi \rightarrow \uptheta)$. Thus, either $t^- : \uppsi \in \mathcal{B}$ or $x^+ : \uptheta \in \mathcal{B}$, for some $t^-, x^+ \in \mathsf{L}_{\mathcal{B}}$. Therefore, by Proposition~\ref{allsets}, either $t \sim t^-$ or $x \sim x^+$, so either $w \sim t^-$ or $v \sim x^+$. Hence, either $w \in \mathsf{ML}_\mathcal{B}^-$ or $v \in \mathsf{ML}_\mathcal{B}^+$, that is either $w \not \in D$  or $v \in D$.
		
	Now, let us assume that $(w \tilde{\rightarrow} v) = u \not \in D$, that is $u \in \mathsf{ML}_\mathcal{B}^-$. Therefore, the rule $(\rightarrow^-)$ applies to $y : (\uppsi \rightarrow \uptheta)$. Thus, $t^+ : \uppsi$ and $x^- : \uptheta$ are on $\mathcal{B}$, for some $t^+, x^- \in \mathsf{L}_{\mathcal{B}}$. Thus, by Proposition~\ref{allsets},  $t \sim t^+$ and $x \sim x^-$, so $w \sim t^+$ and $v \sim x^-$. Hence, $w \in \mathsf{ML}_\mathcal{B}^+$ and $v \in \mathsf{ML}_\mathcal{B}^-$, that is $w \in D$ and $v \not \in D$.\end{proof}

	\subsection{Proof of Proposition~\ref{equiv}}

	\begin{repproposition}{equiv}
		$\tilde{\equiv}$ is a function on $U$ and for all $w, v \in U$ the following holds: 
		\begin{enumerate}
			\item[$(*)$] $w \tilde{\equiv}v \in  D$ iff $w =v$.
		\end{enumerate} 
	\end{repproposition}
	
	\begin{proof} Let $w, v \in U$. In order to prove that $\tilde{\equiv}$ is a function on $U$, we reason in a similar way to the one form the proof of Proposition~\ref{implication}, but instead of $(\equiv^{\rightarrow})$ we use the rule $(\equiv^{\equiv})$. Now, we will show that $\tilde{\equiv}$ satisfies $(*)$. Let $w, v \in U$. Observe that  if $w = \mathbf{w}^+$ or $v = \mathbf{w}^+$ or $(w, v)$ is not $(\equiv)$-closed, then the proof of $(*)$ easily follows  from the definition of $\tilde{\equiv}$. So let $(w, v)$ be $(\equiv)$-closed. Then, $w, v \in \mathsf{ML}_\mathcal{B}$ and there are  $\uppsi, \uptheta \in \mathsf{FOR}$, $u \in \mathsf{ML}_\mathcal{B}$, and $t, x, y  \in \mathsf{L}_{\mathcal{B}}$ such that $w = t$, $v = x$, $u = y$, $t : \uppsi$, $x : \uptheta$, $y : (\uppsi \equiv \uptheta)$ occur on the branch $\mathcal{B}$. 
		
		Assume $(w \tilde{\equiv} v) = u \in D$, that is, $u \in \mathsf{ML}_\mathcal{B}^+$, so $y \in \mathsf{L}^+$. Then, the rule $(\equiv^+)$ applies to $y : (\uppsi \equiv \uptheta)$. Thus, either $t^+:\uppsi$, $x^+: \uptheta$, $t^+=x^+$ are on $\mathcal{B}$ or $t^-:\uppsi$, $x^-: \uptheta$, $t^-=x^-$ belong to $\mathcal{B}$, for some $t^+, t^-, x^+, x^- \in \mathsf{L}_{\mathcal{B}}$. Thus, by Proposition~\ref{allsets}, either $t \sim t^+ \sim x^+ \sim x$ or $t \sim t^- \sim x^- \sim x$, that is, $t \sim x$. Hence, $w \sim v$, that is $w = v$.  
		
		If $(w \tilde{\equiv} v) = u \not \in D$, then $u \in \mathsf{ML}_\mathcal{B}^-$, so $y \in \mathsf{L}^-$. Therefore, the rule $(\equiv^-)$ applies to $y : (\uppsi \equiv \uptheta)$, which means that either of the following cases holds, for some $t^+, t^-, x^+, x^- \in \mathsf{L}_{\mathcal{B}}$:\begin{multicols}{2}
			\begin{enumerate}  
				\item $t^+:\uppsi$, $x^+: \uptheta$, $t^+ \neq x^+$ are on $\mathcal{B}$,
				
				\item $t^+:\uppsi$, $x^-: \uptheta$ are on $\mathcal{B}$,
				
				\item $t^-:\uppsi$, $x^+: \uptheta$ are on $\mathcal{B}$,
				
				\item $t^-:\uppsi$, $x^-: \uptheta$, $t^- \neq x^-$ are on $\mathcal{B}$,
			\end{enumerate}
		\end{multicols}
		
		\noindent By way of example, we will consider cases 1. and 2., as the remaining cases can be proved in an analogous way.  Assume $t^+:\uppsi$, $x^+: \uptheta$, $t^+ \neq x^+$ are on $\mathcal{B}$. By Proposition~\ref{allsets}, we obtain that $t \sim t^+$ and $x \sim x^+$, so $t^+ \sim w$ and $x^+ \sim v$. Observe that $t^+ = x^+$ cannot belong to $\mathcal{B}$, because otherwise both $t^+ = x^+$ and $t^+ \neq x^+$ would belong to $\mathcal{B}$, which by the rule $(\bot_1)$ would imply the closeness of $\mathcal{B}$. Thus, $t^+ \not \sim x^+$, and hence $w \neq v$. Assume that $t^+:\uppsi$, $x^-: \uptheta$ are on $\mathcal{B}$. Then, by Proposition~\ref{allsets}, we obtain that $t \sim t^+$ and $x \sim x^-$, so $t^+ \sim w$ and $x^- \sim v$. Thus, $w \in D$ and $v \not \in D$, which means that $w \neq v$.\end{proof}
	
	\subsection{Proof of Proposition~\ref{valuation}}
	
	\begin{repproposition}{valuation} The function $V$ is well defined and it is a valuation in $\mathcal{M}_{\mathcal{B}}$. \end{repproposition}
	
	\begin{proof} Observe that for all $p \in \mathsf{AF}$ and $w^+, w^- \in \mathsf{L}_{\mathcal{B}}$, it cannot be the case that both $w^+: p$ and $w^- : p$ belong to $\mathcal{B}$; otherwise, $w^+=w^-$ would occur on $\mathcal{B}$, which, by the rule $(\bot_2)$, would mean that $\mathcal{B}$ is closed. Moreover, if $w : p$, $w \sim u$ and $w' : p$, $w' \sim u'$, for some $w,w',u,u' \in \mathsf{L}_{\mathcal{B}}$, then by Proposition~\ref{allsets}, $w \sim w'$, and thus $u \sim u'$. Therefore, $V$ is well defined on $\mathsf{AF}$. Furthermore, since $\tilde{\neg}$, $\tilde(\rightarrow)$, $\tilde(\equiv)$ are functions on $U$, for all $\uppsi, \uptheta, \zeta, \upchi \in \mathsf{FOR}_{\mathcal{B}}$ and $\# \in \{\rightarrow, \equiv\}$ the following holds: 
		\begin{enumerate}[leftmargin=*,labelindent=\parindent]
			\item[(*)] If $V(\uppsi) = V(\uptheta)$, then $V(\neg \uppsi) = V(\neg \uptheta)$
			\item[(**)] If $V(\uppsi) = V(\uptheta)$ and $V(\zeta)= V(\upchi)$, then $V(\uppsi \# \zeta) = V(\uptheta \# \upchi)$.\qedhere
		\end{enumerate}
	\end{proof}

	\subsection{Proof of Proposition~\ref{formulavalue}}

	\begin{repproposition}{formulavalue}
		For all $\uppsi \in \mathsf{FOR}$ and $w \in \mathsf{L}_{\mathcal{B}}$ it holds that:  
		\begin{enumerate}
			\item[$(*)$] If $w: \uppsi \in \mathcal{B}$, then $w \sim V(\uppsi)$.
		\end{enumerate}
	\end{repproposition}
	
	\begin{proof} The proof is by induction on the complexity of formulas. Let $w : p \in \mathcal{B}$, for some $w \in \mathsf{L}_{\mathcal{B}}$. Then, there must exist $u \in  \mathsf{ML}_\mathcal{B}$ such that $w \sim u$. By the definition of $V$, $V(p) = u$, and hence $w \sim V(p)$. Therefore, $(*)$ holds for all $p \in \mathsf{AF}$. Assume that $(*)$ holds for formulas $\uptheta, \upchi$. We will show that it holds for $\neg \uptheta$, $\uptheta \rightarrow \upchi$, $\uptheta \equiv \upchi$.

		Let $w : \neg \uptheta \in \mathcal{B}$, for some $w \in \mathsf{L}_{\mathcal{B}}$, and let $u \in  \mathsf{ML}_\mathcal{B}$ be such that $w \sim u$, that is, $w = u$ is on $\mathcal{B}$. Then, either the rule $(\neg^+)$ or $(\neg^-)$ applies to $w : \neg \uptheta \in \mathcal{B}$, and thus $v : \uptheta$ belongs to $\mathcal{B}$, for some $v \in \mathsf{L}_{\mathcal{B}}$. By the inductive hypothesis, $v \sim V(\uptheta)$, so $v = V(\uptheta)$ occurs on $\mathcal{B}$. Moreover, since $w: \neg \uptheta$, $v : \uptheta$, $w=u$, and $v = V(\uptheta)$ are on $\mathcal{B}$, by the definition of $\tilde{\neg}$, we obtain that $\tilde{\neg} V(\uptheta) = u$. Therefore, by the definition of $V$, $V(\neg \uptheta) = (\tilde{\neg} V(\uptheta)) = u$, for $u \in  \mathsf{ML}_\mathcal{B}$  such that $w \sim u$, which completes the proof of $(*)$ for $\neg \uptheta$. 
		
		Let $w : \uptheta \rightarrow \upchi \in \mathcal{B}$, for some $w \in \mathsf{L}_{\mathcal{B}}$ and let $u \in  \mathsf{ML}_\mathcal{B}$ be such that $(w = u) \in \mathcal{B}$. Then, one of the rules $(\rightarrow^+)$ or $(\rightarrow^-)$ had to be applied to $w : \uptheta \rightarrow \upchi$. Thus, there are $v, t \in \mathsf{L}_{\mathcal{B}}$ such that $v: \uptheta$ and $t: \upchi$ are on $\mathcal{B}$. By the inductive hypothesis, $v \sim V(\uptheta)$ and $t \sim V(\upchi)$, that is $v = V(\uptheta)$ and $t = V(\upchi)$ occur on $\mathcal{B}$. Thus, by the definition of $\tilde{\rightarrow}$, $V(\uptheta) \tilde{\rightarrow} V(\upchi) = u$. On the other hand, by the definition of $V$, we know that $V(\uptheta \rightarrow \upchi) = V(\uptheta) \tilde{\rightarrow} V(\upchi)$, so $V(\uptheta \rightarrow \upchi) = u$, where $w \sim u$. Therefore, $(*)$ holds for $\uptheta \rightarrow \upchi$.
		
		Let $w : \uptheta \equiv \upchi \in \mathcal{B}$, for some $w \in \mathsf{L}_{\mathcal{B}}$ and let $u \in  \mathsf{ML}_\mathcal{B}$ be such that $(w = u) \in \mathcal{B}$. Then, one of the rules $(\equiv^+)$ or $(\equiv^-)$ had to be applied to $w : \uptheta \equiv \upchi$. Thus, there are $v, t \in \mathsf{L}_{\mathcal{B}}$ such that $v: \uptheta$ and $t: \upchi$ are on $\mathcal{B}$. By the induction hypothesis, $v \sim V(\uptheta)$ and $t \sim V(\upchi)$, that is $v = V(\uptheta)$ and $t = V(\upchi)$ occur on $\mathcal{B}$. Thus, by the definition of $\tilde{\equiv}$, $V(\uptheta) \tilde{\equiv} V(\upchi) = u$. On the other hand, by the definition of $V$, we know that $V(\uptheta \equiv \upchi) = V(\uptheta) \tilde{\equiv} V(\upchi)$, so $V(\uptheta \equiv \upchi) = u$, where $w \sim u$. Therefore $(*)$ holds for $\uptheta \equiv \upchi$.\end{proof}
	
	\subsection{Proof of Proposition~\ref{prop::UrfatherProof}}
	
	\begin{repproposition}{prop::UrfatherProof}
		For every $\sf SCI$-formula $\upvarphi$, if $\upvarphi$ has a $\sf TC_{SCI}$-tableau proof, then $\upvarphi$ has $\sf TC_{SCI}+(UB)$-tableau proof. 
	\end{repproposition}
	
	\begin{proof}
		Let $\sf child_\mathcal{B}\in L_\mathcal{B}\times L_\mathcal{B}$, where $\sf L_\mathcal{B}$ is the set of all labels occurring on $\mathcal{B}$. Let two labels $w,v\in\mathsf{L}_\mathcal{B}$ be in the $\mathsf{child}_{\mathcal{B}}$ relation if labelled formulas $w:\upvarphi,v:\uppsi$ are on $\mathcal{B}$ and $v:\uppsi$ appeared on $\mathcal{B}$ as the result of an application of a decomposition rule to $w:\upvarphi$. Let $w,v\in\mathsf{L}_\mathcal{B}$ be in the $\mathsf{descendant}_\mathcal{B}$ relation if they are in the transitive closure of the $\mathsf{child}_\mathcal{B}$ relation. Now, let $\mathsf{L}_\mathcal{B}^\mathsf{d}(w)$ denote all the labels which are in $\mathsf{descendant}_\mathcal{B}$ relation with $w:\upvarphi$. We show that, for each formula $\upvarphi$, whenever $\sf TC_{SCI}$ yields a closed tableau for $\upvarphi$, $\sf TC_{SCI}+(UB)$ yields a closed tableu for $\upvarphi$, too.
		
		Let $\mathcal{B}$ be a branch of a tableau yielded by ${\sf TC}_{\sf SCI}$. Assume that $w:\uppsi$ and $v:\uppsi$ appear on $\mathcal{B}$, $w:\uppsi$ is the $\uppsi$-urfather on $\mathcal{B}$. Of course, $v\notin\mathsf{L}_\mathcal{B}^\mathsf{d}(w)$, for decomposition rules strictly decrease the complexity of formulas they are applied to. Without loss of generality we can assume that $w$ and $v$ have the same polarity, for otherwise $\mathcal{B}$ would get closed by $(\bot_2)$. It suffices to observe that, thanks to the rule $(\mathsf{F})$, there exists a one-to-one mapping $g$ between the sets $\mathsf{L}_\mathcal{B}^\mathsf{d}(w)$ and $\mathsf{L}_\mathcal{B}^\mathsf{d}(v)$ such that $u:\uppsi\in\mathcal{B}$ iff $g(u):\uppsi\in\mathcal{B}$ and $(u,y)\in\mathsf{child}_\mathcal{B}$ iff $(g(u),g(y))\in\mathsf{child}_\mathcal{B}$. By the openness of $\mathcal{B}$, we get that for each $u\in\mathsf{L}_\mathcal{B}^\mathsf{d}(w)$, $u$ and $f(u)$ have the same polarity. Otherwise, after applying ($\sf F$) and ($\bot_2$), $\mathcal{B}$ would get closed. Moreover, for each $u\in\mathsf{L}_\mathcal{B}$ and each $y\in\mathsf{L}_\mathcal{B}^\mathsf{d}(v)$, if $u=y\in\mathcal{B}$, then $u=g^{-1}(y)\in\mathcal{B}$ and no $x\in\mathsf{L}_\mathcal{B}^\mathsf{d}(v)$ is needed to introduce the latter equality to $\mathcal{B}$.
		Indeed, if $u=y$ was introduced to $\mathcal{B}$ by an application of $(\equiv^+)$, it means that there is $x^+\in\mathsf{L}_\mathcal{B}^\mathsf{d}(v)\cup\{v\}$ and $\uppsi,\upchi\in\mathsf{FOR}$ such that $x^+:\uppsi\equiv\upchi$ and $(u,x^+),(v,x^+)\in\mathsf{child}_\mathcal{B}$. By assumption, $g^{-1}(x^+):\uppsi\equiv\upchi\in\mathcal{B}$ or, if $x^+=w$, $w:\uppsi\equiv\upchi$, so $g^{-1}(u)=g^{-1}(y)$.
		
		If the occurrence $u=y$ on $\mathcal{B}$ was a result of an application of $(\sf F)$, then, obviously, there exists $\uppsi\in\mathsf{FOR}$ such that $u:\uppsi,y:\uppsi\in\mathcal{B}$. By the definition of $g$, $g^{-1}(u):\uppsi$, so after applying $(\mathsf{F})$ to $y:\uppsi$ and $g^{-1}(y):\uppsi$, we get $u=g^{-1}(y)\in\mathcal{B}$.
		
		If $u=y$ appeared on $\mathcal{B}$ by an application of $(\equiv^\neg)$, $(\equiv^\to)$ or $(\equiv^\equiv)$, then we employ the induction on the complexity of premises to prove that $u=g^{-1}(y)\in\mathcal{B}$ and no $x\in\mathsf{L}_\mathcal{B}^\mathsf{d}(v)$ was needed to introduce this equality statement to the branch. The reasoning is tedious, but rather straightforward, so we skip the details.
		
		Finally, if an inequality statement $u\neq y$ appears on $\mathcal{B}$ such that $\mathsf{L}_\mathcal{B}^\mathsf{d}(v)$, then it must have been introduced to $\mathcal{B}$ by $(\equiv^-)$. So there is $x^-\in\mathsf{L}_\mathcal{B}^\mathsf{d}(v)\cup\{v\}$ and $\uppsi,\upchi\in\mathsf{FOR}$ such that $x^-:\uppsi\equiv\upchi$ and $(u,x^-),(v,x^-)\in\mathsf{child}_\mathcal{B}$. By assumption, $g^{-1}(x^-):\uppsi\equiv\upchi\in\mathcal{B}$ or, if $x^-=w$, $w:\uppsi\equiv\upchi$, so $g^{-1}(u)\neq g^{-1}(y)$.
		
		From the reasoning above, we can derive that for any equality $w^+=v^-\in\mathcal{B}$, where $v^-\in\mathsf{L}_\mathcal{B}^\mathsf{d}(u)$ for $u:\uppsi\in\mathcal{B}$ which is not the $\uppsi$-urfather on $\mathcal{B}$, there exists an equality $w^+=y^-\in\mathcal{B}$, where $y^-\in\mathsf{L}_\mathcal{B}^\mathsf{d}(x)$ for $x:\uppsi\in\mathcal{B}$ which is the $\uppsi$-urfather on $\mathcal{B}$. The same holds for $w^+$. Likewise, for any pair of expressions $w=v,w\neq v\in\mathcal{B}$, where $v\in\mathsf{L}_\mathcal{B}^\mathsf{d}(u)$ for $u:\uppsi\in\mathcal{B}$ which is not the $\uppsi$-urfather on $\mathcal{B}$, there exists a pair equality $w=y,w\neq y\in\mathcal{B}$, where $y\in\mathsf{L}_\mathcal{B}^\mathsf{d}(x)$ for $x:\uppsi\in\mathcal{B}$ which is the $\uppsi$-urfather on $\mathcal{B}$. The same holds for $w$. It follows that for each branch $\mathcal{B}$ of a $\sf TC_{SCI}$-tableau $\mathcal{T}$, if a closure rule is applicable on $\mathcal{B}$, then it is applicable to a restriction $\mathcal{B}'$ of $\mathcal{B}$ where the condition $\sf(UB)$ is imposed. Thus, for each formula $\upvarphi$, if $\sf TC_{SCI}$ yields a closed tableau for $\upvarphi$, $\sf TC_{SCI}+(UB)$ yields a closed tableau for $\upvarphi$, too.
	\end{proof}
	
	\section{Test formulas}
	In Section~\ref{sect::implementation},
	we denote by~$\upvarphi$ the following formula:
\begin{align*}
(((\pv{q} \equiv \pv{p}) &\rightarrow (\pv{p} \rightarrow \pv{r}))
		\equiv ((\pv{p} \rightarrow (\pv{p} \leftrightarrow \pv{p})) \equiv \pv{p}))\\
&\rightarrow (((\pv{r} \land \pv{p}) \leftrightarrow (\pv{p} \equiv \pv{p}))
		\lor ((\pv{p} \land \pv{p}) \lor \lnot \pv{q}))
\end{align*}

       The transformation~$T$ is reminiscent of the
       well-known polynomial-time reduction
       of general~$\mathsf{SAT}$ to $3\mathsf{SAT}$.
       Given a formula~$\upvarphi$,
       we start by introducing a fresh variable~$\pv{v}_\uppsi$
       for each subformula~$\uppsi$ of~$\upvarphi$.
       Then we recursively define sets~$\sigma(\uppsi)$,
       which intuitively say that $\pv{v}_\uppsi$ represents~$\uppsi$,
       or more formally:
       \begin{eqnarray*}
       \sigma(\pv{p}) &=& \{\pv{v}_{\pv{p}} \equiv \pv{p}\},\\
       \sigma(\neg \uppsi) &=& \sigma(\uppsi)
	       \cup \{\pv{v}_{\neg \uppsi} \equiv \neg \pv{v}_{\uppsi}\},\\
       \sigma(\uppsi \to \uptheta) &=& \sigma(\uppsi) \cup \sigma(\uptheta)
               \cup \{\pv{v}_{\uppsi \to \uptheta}
		       \equiv (\pv{v}_\uppsi \to \pv{v}_\uptheta) \},\\
       \sigma(\uppsi \equiv \uptheta) &=& \sigma(\uppsi) \cup \sigma(\uptheta)
               \cup \{\pv{v}_{\uppsi \equiv \uptheta}
		       \equiv (\pv{v}_\uppsi \equiv \pv{v}_\uptheta) \}.\\
       \end{eqnarray*}
       Now, if $\sigma(\upvarphi) = \{ \uppsi_1, \ldots, \uppsi_n \}$,
       we set
       \[
         T(\upvarphi) = \uppsi_1 \to (\uppsi_2 \to (\ldots \to (\uppsi_n \to \pv{v}_\upvarphi))\ldots)).
       \]
\end{subappendices}

\end{document}